\documentclass[sigplan,10pt]{acmart}
\usepackage{stmaryrd}
\usepackage{tikz}
\usetikzlibrary{shapes,arrows}
\usetikzlibrary{intersections}
\usetikzlibrary{automata}
\usetikzlibrary{positioning}
\usepackage{booktabs}
\usepackage{listings}
\usepackage{xcolor}
\usepackage{multirow}

\newcommand{\OMIT}[1]{}

\lstdefinelanguage{SMTLIB}{
  morekeywords={set-logic, declare-fun, assert, check-sat, get-model},
  sensitive=true,
  morecomment=[l];,
  morestring=[b]"
}
\lstdefinelanguage{Isabelle}{
    keywords={definition,theorem,lemma,proof,qed,by,auto,simp,apply,done,
              where,fixes,assumes,shows,if,then,else,let,in,case,of,
              record,datatype,type_synonym,fun,function,primrec},
    sensitive=true,
    morecomment=[s]{(*}{*)},
    morestring=[b]",
  }

\begin{document}

\setlength{\pdfpageheight}{\paperheight}
\setlength{\pdfpagewidth}{\paperwidth}
\author{Shuanglong Kan}
\affiliation{
  \institution{Barkhausen Institute}
  \country{Germany}
}

\author{Anthony W. Lin}
\affiliation{
  \institution{TU Kaiserslautern \& MPI-SWS}
  \country{Germany}
}

\title{Certified Symbolic Finite Transducers: Formalization and Applications to
String Analysis}

\begin{abstract}
    Finite Transducers (FTs) extend the capabilities of Finite Automata (FAs) by
    enabling the transformation of input strings into output strings. In many
    practical applications --- including program analysis, string constraint
    solving, and analysis of security-critical sanitizers --- Symbolic FTs 
    (SFTs) and Symbolic FAs (SFAs) are used instead of the explicitly
    represented models. To circumvent to the notorious state-space explosion 
    problem caused by an extremely large alphabet size (e.g. Unicode), SFTs
    and SFAs allow the representation of the alphabet as an effective boolean
    algebra including finite unions of intervals, as well as 
    SMT-Algebras.
    %can be instantiated with various Boolean algebras, 
    The
    security-critical nature of many of these applications demands trustworthy
    implementations of such systems. To this end, we present the first 
    formalization of SFTs and their most important algorithms in
    Isabelle/HOL. 
    %This symbolic approach enables not only efficient FT operations—such as computing the output language of an FT given a regular language—but also supports infinite alphabets.
%
    To evaluate the effectiveness of our formalization, we apply the formalized 
    SFTs to two applications: (1) sanitizers for web applications used for 
    preventing XSS attacks, and (2) string solving, which increasingly
    employs intricate string replacement operations.
    %a string solver that models replacement operations with various semantics, including left-most and replace-all matching. 
    Our experimental results demonstrate that our methods are competitive with
    the existing unverified implementations.

    %that our approach achieves computational efficiency in constraint-solving scenarios.

    %, thereby providing a more expressive framework for operations that encompass both recognition and transformation. 
%FTs have wide applications, including program analysis, string constraint 
    %solving, and security-critical sanitizer analysis. 
%However, there is currently no scalable formalization of FTs. This is primarily because transition labels in FTs are not formalized symbolically, resulting in transition explosion and significant performance bottlenecks in practical applications.

    \OMIT{
In this paper, we present a formalization of FTs in the symbolic setting using Isabelle/HOL, where transition labels can be instantiated with various Boolean algebras, such as intervals and arithmetic predicates. This symbolic approach enables not only efficient FT operations—such as computing the output language of an FT given a regular language—but also supports infinite alphabets.
To evaluate the effectiveness of our formalization, we apply the formalized SFTs to two applications: (1) \texttt{HTMLdecode}, a sanitizer used for preventing XSS attacks, and (2) a string solver that models replacement operations with various semantics, including left-most and replace-all matching. Experimental results demonstrate that our approach achieves computational efficiency in constraint-solving scenarios.
    }
\end{abstract}

\maketitle

\section{Introduction}
\label{sec:introduction}

Finite Automata (FA) and Finite Transducers (FT) are fundamental constructs in the theory of formal languages, with extensive applications in programming 
languages and software engineering. Despite this, many practical applications
that employ FAs and FTs --- for example, string analysis for web applications
%character sanitization in web
%applications 
(which easily prone to XSS vulnerabilities) and string constraint solving
--- require extremely large alphabets (e.g. Unicode), which are beyond the
capabilities of classical automata algorithms
\cite{DV21,cav/DAntoniV17,uss/HooimeijerLMSV11,Berkeley-JavaScript}.
More precisely, explicit automata/transducer representations would yield 
an extremely large number of transitions (potentially, one for 
each letter).
%For example, recent advancements in string solvers, as demonstrated by \cite{pacmpl/ChenFHHHKLRW22}, have illustrated the relationship between regular expressions in modern programming languages and various forms of FAs and FTs. Additionally, FAs and FTs find significant industrial applications, such as in the verification of AWS access control policies \cite{DBLP:conf/fmcad/BackesBCDGLRTV18}.

%One significant drawback from classical automata approach is that transition 
%labels are typically non-symbolic and finite. A conventional transition is represented as $q\xrightarrow{a}q'$, where $a$ is a character from a finite alphabet. This simplistic representation can lead to a phenomenon known as transition explosion. 
%For example, if the alphabet $\Sigma$ encompasses the entire Unicode range, which is common in modern programming languages, it consists of $\texttt{0x110000}$ distinct characters. Defining a transition from state $q$ to $q'$ that accepts any character in $\Sigma$ would require splitting into $\texttt{0x110000}$ individual transitions, rendering the FA and FT operations highly inefficient. Furthermore, in practical applications, it is often necessary to consider infinite alphabets, such as the set of all integers.

%Even though there are various formalizations of FAs and FTs in interactive proof assistants such as Isabelle \cite{isabelle-homepage} and Coq \cite{coq-homepage}, these are predominantly based on classical definitions. However, these traditional approaches present certain limitations when applied to practical scenarios. 

Symbolic Finite Automata (SFA) and Symbolic Finite Transducers (SFT)
\cite{cav/DAntoniV17, VeanesHLMB12Transducer,DV21} were proposed around 15 years ago
to deal with the alphabet-explosion problem that often arises in practical
applications. 
%represent advanced extensions of classic FAs and  FTs, enhancing their applicability in practical scenarios.
%
These symbolic extensions permit transition labels defined by an effective 
boolean algebra, allowing for a succinct representation of a set of transitions
that can be conveniently manipulated. 
%more expressive representations. 
For example, in the interval algebra, a transition label might be specified as 
an interval [$\texttt{a}-\texttt{z}$], encompassing all characters from
$\texttt{a}$ to $\texttt{z}$. As another example, in the SMT-Algebra
(particularly, of the theory of Linear Integer Arithmetic), %or 
one could associate an arithmetic condition ($x \mod 2 = 0$) to a transition, 
denoting all even numbers. Efficient algorithms for manipulating (e.g. computing
closures) and analyzing (e.g. checking emptiness) SFA and SFT are by now
rather mature (e.g see the CACM article \cite{DV21}).
SFA and SFT have been applied in numerous domains including analysis of web
applications against XSS vulnerabilities \cite{VeanesHLMB12Transducer,uss/HooimeijerLMSV11},
runtime behavior monitoring \cite{osdi/YaseenABCL20}, automated program
transformation \cite{pldi/HuD17}, and finally SMT solving for the unicode theory
of strings (e.g. see \cite{pacmpl/ChenFHHHKLRW22,CHL+19}).
%Furthermore, numerous studies have demonstrated the broad applicability of SFTs across diverse domains. The work of \cite{VeanesHLMB12Transducer} showcases SFTs in security-critical applications, particularly for cross-site scripting (XSS) prevention. \cite{uss/HooimeijerLMSV11} extends this security focus by applying SFTs to web application sanitizer analysis. In system verification, \cite{osdi/YaseenABCL20} employs SFTs for runtime behavior monitoring, while \cite{pldi/HuD17} demonstrates their utility in automated program transformation through systematic inversion techniques.
%This symbolic approach not only provides a more succinct representation but also supports infinite alphabets, thereby extending the expressive capabilities of FAs and FTs.

Owing to the security-critical nature of many applications of string analysis
(e.g. analysis of web applications against XSS vulnerabilities
\cite{Berkeley-JavaScript} and
at AWS for analyzing Role-Based 
Access Control (RBAC) policies \cite{neha,DBLP:conf/fmcad/BackesBCDGLRTV18}), 
there is an 
increasing need of trustworthy implementations in such application domains. 
Unfortunately, string analysis implementations are often quite buggy. For
example, even state-of-the-art SMT solvers over the theory of strings 
were found to be buggy through such techniques as fuzzing
\cite{DBLP:conf/cav/BlotskyMBZKG18,BM20,Mansur20}. 
%Recent applications of string solvers in security critical applications
%(e.g. at AWS for analyzing Role-Based 
%Access Control (RBAC) policies \cite{neha,DBLP:conf/fmcad/BackesBCDGLRTV18})
This
has motivated initial formalization efforts for the theory of strings in 
Isabelle/HOL, e.g., a verified solver \cite{cpp/KanLRS22}
for handling a small fragment of SMT-LIB 2.6 unicode theory of strings
consisting only of equality, concatenation and regular constraints, and a
formalization of the semantics of this theory \cite{verified-verifying}.

Despite these, existing verified implementations for string analysis do not
handle the following indispensable string operations for analysis of web
applications (e.g. see
\cite{DBLP:conf/popl/LinB16,Kern,Berkeley-JavaScript,systematic-transduction,uss/HooimeijerLMSV11}):
\emph{string transductions} and \emph{replaceall} operators. In particular, 
string transductions are used frequently in web applications (e.g. for
sanitization, as well as implicitly applied transductions, for instance
through \texttt{innerHTML}) and are well-known to be critical for analyzing XSS
vulnerabilities for web applications
\cite{Berkeley-JavaScript,systematic-transduction,Kern,uss/HooimeijerLMSV11,DBLP:conf/popl/LinB16}. In particular, SMT-LIB string constraints that 
arise from such applications require the \emph{replaceall} operator, which is currently not handled by any existing verified
implementation. This paper tackles the formalization of SFA/SFT and
their associated algorithms, as well as applications to analysis of sanitizers
and verified solver for string constraints. 

\paragraph{Contributions.} 
In this work, we present a comprehensive formalization of SFTs. 
%However, the formalization of transition labels in SFAs and SFTs presents significant challenges within interactive proof assistants. 
Two fundamental considerations arise in such a formalization: first, the 
representation of transition labels must be sufficiently expressive to accommodate diverse predicate representations of boolean algebras; second, the formalization framework must be designed with extensibility in mind to facilitate the incorporation of new boolean algebras while minimizing redundant proof efforts.

To address these challenges, %in supporting diverse transition label theories, 
we adopt a refinement-based approach. At the abstraction level, transition labels are formalized through the fundamental mathematical concept of \emph{sets}. This abstraction facilitates subsequent refinement to various representations of Boolean algebras, such as intervals and arithmetic predicates, while maintaining theoretical consistency.

The key operation of our formalization is the product operation between an SFT and an input regular language. Specifically, given an SFA $\mathcal{A}$ representing a regular language and an SFT $\mathcal{T}$, we define the product operation $\mathcal{T} \times\mathcal{A}$ that characterizes the output language generated by $\mathcal{T}$ when processing inputs from the language recognized by $\mathcal{A}$.

In the refinement level, we implemented transition labels using an interval-based representation to examplify the refinement process. The formalization of interval algebra provides efficient set-theoretic operations—including membership checking, intersection, and difference computations—facilitating the refinement of transition labels from abstract sets to concrete intervals. Furthermore, leveraging the data refinement framework \cite{DBLP:conf/itp/Lammich13} in Isabelle/HOL, we store states and transitions using sophisticated data structures such as hashmaps and red-black trees, ensuring efficient automata manipulation.

To evaluate the effectiveness and efficiency of our formalization, we 
provide two applications: (1) analysis of the web application sanitizers for html, css, json, and javascript
which are used for preventing XSS attacks, and (2) string solving handling equality, concatenation, regular
constraints, and more importantly the string replaceall operator, which has been
argued to be indispensable for analysis of XSS
\cite{DBLP:conf/popl/LinB16,Kern,Berkeley-JavaScript,systematic-transduction,uss/HooimeijerLMSV11}.
The experimental results demonstrate that our formalization achieves computational efficiency in constraint-solving scenarios.

\OMIT{
which increasingly
extended

the SMT-LIB \cite{smtlib} compliant certified string solver \cite{cpp/KanLRS22}
with replacement operations. We evaluated our implementation using a set of benchmarks from the SMT-LIB repository \cite{smtlib_benchmarks}.
}

In summary, our formalization makes the following contributions:
\begin{enumerate}
\item We present the first formalization of symbolic finite transducers  in Isabelle/HOL.
We prove the correctness of key operations of SFTs and symbolic SFAs, such as the product operation between SFTs and SFAs.
\item We leverage the refinement framework to develop an extensible and efficient SFT implementation capable of supporting diverse Boolean algebras, ensuring adaptability to future transition label representations. The correctness of the refined, efficient implementation is proved via structural isomorphism between refinement layers.
\item We develop a certified interval algebra with verified set-theoretic operations, enabling refinement of transition labels from abstract sets to concrete intervals.
\item We demonstrate the practical utility of our formalization by applying it to sanitizer analysis and integrating our formalized SFT into the verified forward-propagation string solver CertiStr \cite{cpp/KanLRS22}. This integration extends CertiStr from supporting only simple concatenation and regular constraints to handling more sophisticated string replacement operations.
\end{enumerate}

\paragraph{Organization.}
The remainder of this paper is organized as follows:
Section \ref{sec:motivation} provides concrete motivating examples.
Section \ref{sec:sft} introduces the definition of symbolic finite transducers and their product operations.
Section \ref{sec:formalization} presents our formalization of SFTs in Isabelle/HOL.
Section \ref{sec:application} demonstrates applications to sanitizer analysis and string constraint solving.
Section \ref{sec:related-work} discusses related work.
Section \ref{sec:conclusion} concludes and outlines future directions.

\section{Motivating Example}
\label{sec:motivation}

In this section, we recall a simple XSS example (see Listing
\ref{lst-web-security}) from the CACM article 
\cite{Kern}, which epitomizes the indispensability of the replace-all operator 
and string transductions, as well as other string operators like concatenation.
In particular, string transductions are so common in web applications that they
sometimes even happen silently (see \texttt{innerHTML} below). A more detailed
account on this can be found in \cite{systematic-transduction}.

\begin{lstlisting}[language={}, caption={Mutation XSS attack code example}, label={lst-web-security}, float=htbp]
var x = goog.string.htmlEscape(cat);
var y = goog.string.escapeString(x);
catElem.innerHTML = '<button onclick=
"createCatList(\'' + y + '\')">' + x + '</button>';
\end{lstlisting}

The code assigns an HTML markup for a button to the DOM element \texttt{catElem}. This button creates a category named \texttt{cat}, whose value is provided by an untrusted third party. To mitigate potential security risks, the code attempts to sanitize the value of \texttt{cat} using the Closure Library~\cite{google-closure-library} string functions \texttt{htmlEscape} and \texttt{escapeString} (the latter implements JavaScript escaping). 

When the input value \texttt{Flora \& Fauna} is provided for \texttt{cat}, the resulting HTML markup is as intended:
\begin{verbatim}
<button onclick="createCatList('Flora \&amp; Fauna
')"> Flora \&amp; Fauna</button>
\end{verbatim}
On the other hand, inputting the value ');alert(1);// to cat,
results in the HTML markup:
\begin{verbatim}
  <button onclick="createCatList('&#39;);alert(1);
  //')"> &#39;);alert(1);//')</button>
\end{verbatim}

When this markup is inserted into the DOM via \texttt{innerHTML}, 
the resulting string appears as \verb|"createCatList('');alert| (\verb|1);..."|. It enables the attacker's script \texttt{alert(1)} to execute immediately after \texttt{createCatList}. 
This subtle XSS vulnerability does not stem from the absence of escaping functions, but rather from applying them in the incorrect order.

Both \texttt{htmlEscape} and \texttt{escapeString} can be naturally modeled as SFTs, since they transform input strings into output strings. The above code snippet can be represented by the following string constraints (where \texttt{w1},
\texttt{w2}, and \texttt{w3} are constant strings):

\begin{verbatim}
  1. x = SFT_htmlEscape(cat)
  2. y = SFT_escapeString(x)
  3. z = w1 + y + w2 + x + w3;
  4. catElem.innerHTML = SFT_innerHTML(z)
  5. catElem.innerHTML matches R_unsafe
\end{verbatim}
\texttt{SFT\_innerHTML} is a transducer implementing the browser transductions upon innerHTML
assignments. \texttt{R\_unsafe} is a regular expression defining unsafe HTML patterns that can lead to XSS vulnerabilities.

In the remainder of this paper, we formalize SFTs and demonstrate how they can be used to verify the correctness of sanitizer functions. We also show how to integrate SFT-based reasoning into existing string solvers, enabling support not only for basic regular and concatenation constraints, but also for more complex string verification problems involving replacement operations.

\section{Symbolic Finite Transducers}
\label{sec:sft}

We recall the definition of SFTs \cite{VeanesHLMB12Transducer}, 
abstracting from specific implementation details in Isabelle/HOL.

Let $\mathcal{U}$ be a multi-sorted carrier set or background universe, which is equipped with functions and relations over the elements. We use $\tau$ as a sort and $\mathcal{U}^\tau$ denotes the sub-universe of elements of type $\tau$.
We have a special type $\mathbb{B}$ with $\mathcal{U}^\mathbb{B} = \{ \top, \bot\}$, which corresponds to the boolean type.

A lambda term is defined as $\lambda x.~t$ of type $\tau_1 \rightarrow \tau_2$.
When $\tau_2$ is $\mathbb{B}$, this lambda term is a predicate. Let $\phi$ be a predicate. We write $a\in \llbracket\phi \rrbracket$ if $\phi~a=\top$. For non-predicate lambda terms, we view them as functions that generate output elements of type $\tau_2$ given input terms of type $\tau_1$.
With these notations and the above definitions, we can define SFTs as follows.

\begin{definition}[Symbolic Finite Transducer]
\label{def-sft}
   A Symbolic Finite Transducer over $\tau_1\rightarrow \tau_2$ is a quadruple $\mathcal{T} = (\mathcal{Q}, \Delta, \mathcal{I}, \mathcal{F})$, where 
   \begin{itemize}
   \item $\mathcal{Q}$ is a finite set of states,
   \item $\mathcal{I}\subseteq \mathcal{Q}$ is the set of initial states,
   \item $\mathcal{F} \subseteq\mathcal{Q}$ is the set of accepting states,
   \item $\Delta$ is the set of transition relations. Each element in $\Delta$ is of the form $(q, \phi, f, q')$ or written as $q\xrightarrow{\phi, f} q'$, where $q$ and $q'$ are states in $\mathcal{Q}$.
   $\phi$ is a predicate of type $\tau_1\rightarrow \mathbb{B}$.
   $f$ is a lambda term of type $\tau_1\rightarrow \tau_2$. $f$ is called an \emph{output function}.
   \end{itemize}

For each transition $q\xrightarrow{\phi, f} q'$, if there exists an element $a\in \llbracket \phi \rrbracket$, where $a$ is called an input,  then the application $(f~a)$ is the output.
   
\end{definition}

SFTs accept an input word and generate an output word. This can be defined by \emph{runs} of SFTs.
An SFT run $\sigma$ is a sequence $(q_0, \phi_0, f_0, q_1),(q_1, \phi_1, f_1, q_2),\ldots, (q_{n-1}, \phi_{n-1}, f_{n-1}, q_n)$ such that $q_0\in \mathcal{I}$ and $(q_i,
\phi_i, f_i, q_{i+1}), 0 \leq i \leq n-1$ is a transition in $\Delta$.
$\sigma$ is \emph{accepting} when $q_n \in \mathcal{F}$.

For a word $w = a_0,\ldots, a_{n-1}$, it is accepted by $\sigma$ if and only if $a_i \in \llbracket\phi_i \rrbracket$ for $0 \leq i \leq n - 1$. When $w$ is accepted, run $\sigma$ generates an output sequence $w'= f_0~a_0, \ldots, f_{n-1}~a_{n-1}$. We define:
\begin{itemize}
\item  $(a_0,(f_0~a_0)), \ldots, (a_{n-1},(f_{n-1}~a_{n-1}))$ a \emph{trace},
\item $a_0,\ldots,a_{n-1}$ the \emph{input} of the trace and $(f_0~a_0), \ldots, $ $(f_{n-1}~a_{n-1})$ the \emph{output} of the trace.
\end{itemize}

If $a_0,\ldots,a_{n-1}$ is accepted by an accepting run in $\mathcal{T}$, we say that the trace is an accepting trace of $\mathcal{T}$.
For a trace $\pi$, we denote its input as $\mathit{in}(\pi)$ and its output as $\mathit{out}(\pi)$.
Given an SFT $\mathcal{T}$ and a word $w$, we define the \emph{product} operation of $\mathcal{T}$ and $w$ (denoted as $\mathcal{T}\times\{w\}$) as the set of outputs generated by $\mathcal{T}$ with input $w$. More precisely,
\[
\begin{split}
\mathcal{T}\times\{w\} = \{w'\mid \exists \pi.~\pi \text{ is an accepting trace of } \mathcal{T} \land \\ \mathit{in}(\pi) = w \land \mathit{out}(\pi) = w'\}.
\end{split}
\]

To make the operation \emph{product} more general, we extend the operation to an SFT and a set of input words represented by a regular language, which can be denoted by an SFA $\mathcal{A}$. More precisely,
\[
\begin{split}
\mathcal{T}\times \mathcal{A} = \{w'\mid \exists w.~w\in \mathcal{L}(\mathcal{A})\land w' \in \mathcal{T}\times\{w\}\}, \\\text{ where }\mathcal{L}(\mathcal{A})\text{ denotes the language of }\mathcal{A}.
\end{split}
\]

\begin{figure}[hbt!]
  \centering
  \begin{tikzpicture}[shorten >=1pt, node distance=2.5cm and 2cm, on grid, auto]
    % Make q0 and q1 horizontally wide apart
    \node[state, initial] (q_0)   {$q_0$}; 
    \node[state] (q_1) [right=4cm of q_0] {$q_1$}; 
    \node[state, accepting] (q_2) [below=2.2cm of q_1] {$q_2$}; 
 
    \path[->] 
    (q_0) edge node {[$\texttt{a}-\texttt{z}$],$f_1$} (q_1)
    (q_1) edge node[right] {[$\texttt{A}-\texttt{Z}$],$f_2$} (q_2)
    (q_1) edge[loop above] node {[$\texttt{a}-\texttt{z}$],$f_1$} (q_1)
    (q_2) edge[loop below] node {[$\texttt{A}-\texttt{Z}$],$f_2$} (q_2);
  \end{tikzpicture}
    \caption{An example of SFT}
    \label{fig-example-ft}
\end{figure}

Figure \ref{fig-example-ft} illustrates an SFT that accepts words matching the regular expression $/[\texttt{a}-\texttt{z}]+[\texttt{A}-\texttt{Z}]+/$. The transition labels utilize intervals of the form $[i\text{-}j]$, which are interpreted as predicates $\lambda x.~i \leq x \leq j$. The output functions $f_1$ and $f_2$ perform case transformations: $f_1 = \lambda x.~\texttt{toUpper}(x)$ converts lowercase letters to uppercase, while $f_2 = \lambda x.~\texttt{toLower}(x)$ performs the inverse operation. For example, given the input string \texttt{bigSMALL}, this SFT produces the output \texttt{BIGsmall}.

%The transition labels can be more complex. For instance, it can be a predicate over integers as: $\lambda x.~ x >0 \land x \text{ mod } 2 = 0$. This symbolic label corresponds to the set of even natural numbers.

\section{The Isabelle/HOL Formalization of SFTs}
\label{sec:formalization}

In this section, we present our formalization of SFTs. We begin by outlining the basic architecture of our Isabelle/HOL formalization, which is organized into three distinct layers.

The \textbf{abstract layer} represents transition labels as \textbf{sets}, with states likewise stored in sets. This abstraction, which deliberately omits implementation details, closely aligns with the SFT definitions presented in Section~\ref{sec:sft}.

The \textbf{implementation layer} refines the abstract transition labels to a locale where a boolean algebra is defined as type variable $'\texttt{b}$ together with a collection of operations and assumptions. State sets are refined using Refine\_Monadic \cite{Refine_Monadic-AFP}, a framework in Isabelle/HOL for data refinement that automatically transforms general sets into efficient data structures such as hashmaps and red-black trees.

The \textbf{boolean algebra layer} refines the transition label locale to a concrete boolean algebra implementation. In this paper, we present an interval-based refinement.

This three-layer architecture separates the fixed components (abstract and implementation layers) of the SFT formalization from the variable components (transition labels). This separation enables us to extend the SFT formalization to accommodate new transition label theories without modifying the first two layer formalization.

\subsection{The Abstract Layer of SFTs}
\label{sec:abstract-layer}

We present the SFT formalization in abstract layer in Fig. \ref{fig-def-FT}. While the elements $\mathcal{Q}_t$, $\mathcal{I}_t$, and $\mathcal{F}_t$ directly correspond to their counterparts ($\mathcal{Q}$, $\mathcal{I}$, and $\mathcal{F}$) in Definition \ref{def-sft}, the transition relations are not exactly the same.
The transition relations are formalized through \texttt{LTTS} (Labeled Transducer Transition System), where each transition is represented as a triple $'q \times ('a \texttt{ set option }\times 'i) \times 'q$. This representation reflects several key design decisions aimed at enhancing the abstraction and flexibility of our SFT formalization.
\begin{figure}[hbt!]
	\begin{lstlisting}
record (|$'q,$||$~'a,$| |$'i,$| |$'b$|) |$\texttt{NFT}$| =
	|$\mathcal{Q}_t$| :: "|$'q$| set"
	|$\Delta_t$| :: "(|$'q,$||$~'a,$| |$'i$|) LTTS"
	|$\mathcal{I}_t$| :: "|$'q$| set"
	|$\mathcal{F}_t$| :: "|$'q$| set"
        |$\mathcal{M}_t$| :: "|$'i\Rightarrow ('a,~'b)$| Tlabel"
type_synonym |$('q,~'a,~'i)$| LTTS = "|$('q \times ('a \text{ set option }\times 'i) \times 'q)$| set"
type_synonym |$('a,~'b)$| Tlabel = "|$'a \text{ option}~\Rightarrow~'b$ set option"
	\end{lstlisting}
\caption{The formalization of SFTs in Isabelle/HOL}
\label{fig-def-FT}
\end{figure}

Firstly, $'a \text{ set option }$ is the input type of the transition, it accepts a set of elements of type $'a$ or $\texttt{None}$ corresponding to empty string $\varepsilon$. 
Accepting a set of $'a$ elements aims to express the same but more abstract semantics of the input labels in Definition \ref{def-sft}, in which an input label is a predicate. A predicate's semantics as introduced before represents a set of elements that make the predicate true. But predicates have various different forms. For instance, the interval $[1-9]$ represents the set $\{e \mid 1 \leq e \leq 9\}$. The predicate $\lambda b.~ b[7] = 1$, where $b$ is a bit vector of length 8, denotes the set of bit vectors that have a 1 in the 7th position. All different forms of labels are abstracted as sets in our formalization.
The value $\texttt{None}$ represents the empty string $\varepsilon$ in our formalization, indicating a transition that consumes no input but may still produce output elements. This design choice facilitates the modeling of real-world applications in SFTs, as we will demonstrate in our application to string solvers.

The second element of type $'i$ in LTTS serves as an index into the output function space. 
We use indices instead of functions %themselves 
to enable the reuse of the same output functions for different transitions.
The mapping $\mathcal{M}_t$ associates each index with a specific output
function. These output functions, formalized by \texttt{Tlabel}, map a single
input element to a set of possible output elements rather than to a single
element. This design enables nondeterministic output behavior, where the
transducer may select any element from the output set arbitrarily or according
to specified criteria. Further, output functions can produce empty string $\varepsilon$ by returning $\texttt{None}$, providing further flexibility in transition behavior.

%Now let us look at an example. In some modern programming languages, there are support for replacement operations : $\texttt{replace}(\texttt{str}, \texttt{pattern}, \texttt{replacement})$. This operation replaces the first occurrence of the substring  in \texttt{str} that matches \texttt{pattern} (which usually a regular expression) with the string \texttt{replacement}). 
%We can easily model this operation using our formalization of NFTs with the following steps:
%\begin{itemize}
%\item Step 1, construct a corresponding NFA from \texttt{pattern}).
%\item Step 2, convert each transition label for transducers by adding a function that maps each input character to empty, that is, generate nothing.
%\item Step 3, Add a new accepting state, and redirect existing accepting states to the new accepting state and transition label as $(\texttt{None}, \lambda x. \texttt{replacement})$.
%\item Step 4, unlabel previous accepting states as non-accepting states. 
%\end{itemize}

%We can see that the definition of transition labels in NFTs makes it is very easy to model the replacement operation.

\paragraph{The product operation formalization.} We formalize the product operation between an SFT $\mathcal{T}$ and an SFA $\mathcal{A}$, denoted as $\mathcal{T} \times \mathcal{A}$ in Section \ref{sec:sft}. The SFA is formalized as a record type \texttt{NFA in Figure \ref{fig-def-FAs}}.
An additional consideration in this operation is the presence of $\varepsilon$-transitions in our SFT formalization, which implies that the resulting automata may also contain $\varepsilon$-transitions.
Therefore, we need to formalize $\varepsilon$SFAs, which is shown in Fig. \ref{fig-def-FAs} as \texttt{eNFA}. The $\varepsilon$SFA formalization extends the standard SFA structure by introducing $\Delta_e'$, which captures $\varepsilon$-transitions as pairs of states, while maintaining the same labeled transition relation $\Delta$ as in standard SFAs.
In this paper, we present only the definitions of  $\varepsilon$SFAs and SFAs, while the complete formalization, including correctness proofs, is available in our Isabelle development.

\begin{figure}[hbt!]
	\begin{lstlisting}
record (|$'q,$||$~'a$|) |$\texttt{NFA}$| =
	|$\mathcal{Q}$| :: "|$'q$| set"
	|$\Delta$| :: "(|$'q,$||$~'a$|) LTS"
	|$\mathcal{I}$| :: "|$'q$| set"
	|$\mathcal{F}$| :: "|$'q$| set"

record (|$'q,$||$~'a$|) |$\texttt{eNFA}$| =
	|$\mathcal{Q}_e$| :: "|$'q$| set"
	|$\Delta_e$| :: "(|$'q,$||$~'a$|) LTS"
	|$\Delta_e'$| :: "|$('q * 'q)$| set"
	|$\mathcal{I}_e$| :: "|$'q$| set"
	|$\mathcal{F}_e$| :: "|$'q$| set"

type_synonym |$('q, 'a)$| LTS = "|$'q \times 'a \text{ set }\times 'q$|"    
	\end{lstlisting}
\caption{The formalization of $\varepsilon$SFAs and SFAs}
\label{fig-def-FAs}
\end{figure}

Having established these foundational definitions, we can now formalize the product operation. 

\begin{figure}[hbt!]
	\begin{lstlisting}
definition productT :: |$"('q,'a,'i,'b)"$| NFT |$\Rightarrow$| 
   |$('q,'a)$| NFA |$\Rightarrow$||$(('a,'b)$| Tlabel |$\Rightarrow 'a \text{ set}$| 
   |$\Rightarrow 'b\text{ set option})\Rightarrow$| |$('q\times 'q, ~'b)$| eNFA where
"productT |$\mathcal{T}$| |$\mathcal{A}$| F = |$\llparenthesis$|
  |$\mathcal{Q}e = \mathcal{Q}_t~\mathcal{T} \times \mathcal{Q} ~\mathcal{A}$|,
  |$\Delta_e = \{((p,p'),~\texttt{the }(((\mathcal{M}_t~\mathcal{T})~f)\texttt{ None}),~(q, p'))\mid $|
       |$p, p', q, f.~p'\in \mathcal{Q}~\mathcal{A}~\land$| |$(p, (\texttt{None}, f), q)\in \Delta_t~\mathcal{T}$|
        |$ \land~\exists S.~(\mathcal{M}_t~\mathcal{T})~f~\texttt{None} = \text{Some } S\} ~\cup$|
      |$\{((p,p'),~\texttt{the }(F~((\mathcal{M}_t~\mathcal{T})~f)~(\sigma_1\cap\sigma_2)),(q, q'))\mid $|
       |$p, p', q, \sigma_1, \sigma_2, q', f.~(p, (\text{Some }\sigma_1, f), q)\in \Delta_t~\mathcal{T} \land$|
       |$ (p', \sigma_2, q')\in\Delta~\mathcal{A}\land \sigma_1\cap \sigma_2 \neq \emptyset~\land$|
          |$ \exists S.~F~((\mathcal{M}_t~\mathcal{T})~f)~(\sigma_1\cap\sigma_2) = \text{Some } S\}$|,
  |$\Delta_e' = \{((p,p'),~(q, p'))\mid p, p', q, f.~p'\in \mathcal{Q}~\mathcal{A} \land$|
       |$(p, (None, f), q)\in \Delta_t~\mathcal{T}~\land $|
       |$(\mathcal{M}_t~\mathcal{T})~f~\texttt{None} = \text{None}\} ~\cup$|
      |$\{((p,p'),~(q, q'))\mid p, p', q, \sigma_1, \sigma_2, q', f.~$|
        |$ (p, (\text{Some }\sigma_1, f), q)\in \Delta_t~\mathcal{T} \land (p', \sigma_2, q')\in\Delta~\mathcal{A}$|
        |$\land~\sigma_1\cap \sigma_2 \neq \emptyset~\land\exists x\in(\sigma_1\cap\sigma_2).$|
        |$~((\mathcal{M}_t~\mathcal{T})~f)~(\text{Some } x) = \text{None} \}$|,
  |$\mathcal{I}_e = \mathcal{I}_t~\mathcal{T} \times \mathcal{I}~\mathcal{A}$|
  |$\mathcal{F}_e = \mathcal{F}_t~\mathcal{T} \times \mathcal{F}~\mathcal{A}$| |$\rrparenthesis$|"
	\end{lstlisting}
\caption{The formalization of product operation}
\label{fig-def-FTProd}
\end{figure}

Figure \ref{fig-def-FTProd} depicts the abstract level formalization for the product of an SFT and an SFA. The parameters $\mathcal{T}$ and $\mathcal{A}$ are an SFT and an SFA, respectively. The 
result of the product operation is an $\varepsilon$SFA.
But we need to explain the role of parameter $\text{\texttt{F}}$. The output function $f$ for each transition in $\Delta_t$ is of type "$'a\;\text{option} \Rightarrow 'b\;\text{set option}$", which applies to a \emph{single} element of type $'a$ or $\varepsilon$. $\text{\texttt{F}}$ extends $f$ to apply $f$ to a set of elements. More precisely, let $f$ be an output function, the semantics of $\text{\texttt{F}}$ is defined as follows:

\[\texttt{F}~f~A=\texttt{Some } (\bigcup_{a\in A} (\text{if }f~a= \text{Some }S \texttt{ then } S \texttt{ else } \emptyset))\].

The transition relations $\Delta_e$ and $\Delta_e'$ are determinted by considering two distinct cases based on the nature of transitions in the SFT: $\varepsilon$-transitions and non-$\varepsilon$-transitions. In both cases, the transition labels in the resulting $\varepsilon$SFA are derived from the composition of the SFT's output function and the SFA's input labels. Let us consider $\Delta_e$ first.

\begin{enumerate}
\item When $(p, (\text{None}, f), q)$ is a transition in $\Delta_t$, i.e. the input character is $\varepsilon$, and $f~\text{None} \neq \text{None}$. Consequently, the SFA $\mathcal{A}$ remains in its current state, and the product transition produces the output "$((\mathcal{M}_t~\mathcal{T})~f)~\text{None}$". Remember that $f$ is just an index, $(\mathcal{M}_t~\mathcal{T})~f$ is the output function.

\item When $(p, (\text{Some}~\sigma_1, f), q)$ is a transition in $\Delta_t$, synchronization is possible only with SFA transitions that share characters with $\sigma_1$, i.e., $\sigma_1\cap\sigma_2\neq \emptyset$, where $\sigma_2$ represents the input label of the corresponding SFA transition. The resulting output is $\texttt{F}~((\mathcal{M}_t~\mathcal{T})~f)~(\sigma_1\cap\sigma_2)$.
\end{enumerate}

The transitions in $\Delta_e'$ follow a similar pattern with analogous cases for $\varepsilon$ and non-$\varepsilon$ transitions.

To establish the correctness specification of the product operation, we begin by formalizing the concept of SFT \emph{traces} as introduced in Definition \ref{def-sft}. In our formalization, traces are represented by the type $('a\;\text{option} \times 'b\;\text{option})\;\texttt{list}$, as shown in Figure \ref{fig-def-output}.
Given a trace $\pi$, we define two key projection functions:
(1) \texttt{inputE}, which corresponds to ${in}(\pi)$ and extracts the input sequence
(2) \texttt{outputE}, which corresponds to ${out}(\pi)$ and extracts the output sequence.

The definition \texttt{outputL} generalizes \texttt{outputE} to characterize the set of all possible outputs that an SFT $\mathcal{T}$ can generate when processing inputs from the language accepted by the SFA $\mathcal{A}$. The reachability of a trace $\pi$ between states $q$ and $q'$ is checked by the predicate $\texttt{LTTS\_reachable}~\mathcal{T}~q~\pi~q'$.

\begin{figure}[hbt!]
	\begin{lstlisting}
fun inputE :: 
   "|$('a\;\text{option}\;\times\;'b\;\text{option})\;\text{list}\;\Rightarrow\;'a\;\text{list}$|"
where
  "inputE [] = []" |$\mid$|
  "inputE ((Some a, |$\_$|) |$\#$| l) = 
                      a |$\#$| (inputE l)" |$\mid$|
  "inputE ((None, |$\_$|) |$\#$| l) = (inputE l)"

fun outputE :: 
    "|$('a \text{ option} \times 'b \text{ option}) \text{ list} \Rightarrow 'b \text{ list}$|" 
where
  "outputE [] = []" |$\mid$|
  "outputE ((_,Some a) # l) = 
                    a # (outputE l)" |$\mid$|
  "outputE ((_,None) # l) = (outputE l)"

definition outputL :: 
   "|$('q, 'a, 'i, 'b)\text{  NFT} \Rightarrow 
                               ('q, 'a) \text{ NFA} \Rightarrow 'b\text{ list set}$|" where
  "outputL |$\mathcal{T}$| |$\mathcal{A}$| = {outputE |$\pi~\mid~\pi~q~q'$|. 
    |$q\in \mathcal{I}_t~\mathcal{T} \land q' \in \mathcal{F}_t~\mathcal{T}~\land$| 
    |$ \text{LTTS\_reachable }\mathcal{T}~q~\pi~q'\land$||$ \text{inputE }\pi\in \mathcal{L}~\mathcal{A}$|}"
	\end{lstlisting}
\caption{The formalization of traces in SFTs}
\label{fig-def-output}
\end{figure}

\begin{figure}[hbt!]
	\begin{lstlisting}
lemma productT_correct:
  fixes |$\mathcal{T}$| |$\mathcal{A}$| F
  assumes 
   F_ok1: "|$\forall~f~s.~(\forall e \in s.~f~(\text{Some } e) = \text{None})\longleftrightarrow$|
       |$\text{ F}~f~s = \text{None}"$|
   and F_ok2: "|$\forall~f~s.~ \texttt{F}~f~s\neq \texttt{None} \longrightarrow$| F |$f~s$| = 
        |$\text{Some }(\cup~\{\texttt{S} \mid e~\texttt{S}.~e \in s~\land~$|
        |$f~(\text{Some }e) = \text{ Some S}\})$|"
      and wfTA: "NFT_wf |$\mathcal{T}~\land \text{ NFA } \mathcal{A}$|"
  |$\textcolor{blue}{\text{shows}}$| "|$\mathcal{L}_e$| (productT |$\mathcal{T}~\mathcal{A}$| F) = outputL 
          |$\mathcal{T}~\mathcal{A}$|"

	\end{lstlisting}
\caption{The correctness lemma of the product operation}
\label{fig-def-product-correct}
\end{figure}

Figure \ref{fig-def-product-correct} presents Lemma \texttt{productT\_correct}, which establishes the correctness of the product operation. The lemma's assumptions, \texttt{F\_ok1} and \texttt{F\_ok2}, specify the essential properties of function \texttt{F}. 
The assumptions $\texttt{NFT\_wf}~\mathcal{T}$ and $\texttt{NFA}~\mathcal{A}$ ensure that the SFT $\mathcal{T}$ and the SFA $\mathcal{A}$ are well-formed.
The conclusion, marked by \textcolor{blue}{\texttt{shows}}, demonstrates that the language of the constructed $\varepsilon$SFA ($\mathcal{L}_e$ denotes the language of $\varepsilon$SFA) from $\mathcal{T} \times \mathcal{A}$ coincides with the mathematical semantics defined by \texttt{outputL}, thereby establishing semantic preservation of the product construction.

\subsection{Implementation Layer Refinement}
\label{sec_alg_refinement}
Having introduced the abstract definition of the SFT product operation in Section \ref{sec:abstract-layer}, we now turn to its algorithmic refinement. In this section, we present an efficient implementation of the product construction and refine the representation of transition labels into a Boolean algebra locale, which serves as the interface structure.

\paragraph{Boolean Algebra Locale}

Figure \ref{fig-bool-algebra-locale} presents the simplified definition of a boolean algebra locale. The type variable \texttt{$'$b} represents any boolean algebra, while the type variable \texttt{$'$a} denotes the element type of the sets that the boolean algebra represents. The locale defines a collection of operations and assumptions that characterize boolean algebra behavior, including operations for set semantics, emptiness checks, non-emptiness checks, intersection, difference, and element membership. The function "\texttt{sem}" provides the semantic interpretation of a boolean algebra as a set, with all other operations' assumptions defined in terms of this set semantics.

The locale serves as an interface to boolean algebra implementations. Any concrete boolean algebra implementation must provide a concrete instantiation of this locale, which includes defining the type variable $'$\texttt{b} and providing implementations for all operations along with proofs of the required assumptions.

\begin{figure}[hbt!]
\begin{lstlisting}[mathescape=true]
locale bool_algebra =
 fixes sem :: " 'b $\Rightarrow$ 'a::ord set"
 fixes empty:: "'b $\Rightarrow$ bool"
 fixes nempty:: "'b $\Rightarrow$ bool"
 fixes intersect:: "'b $\Rightarrow$ 'b $\Rightarrow$ 'b"
 fixes diff:: "'b $\Rightarrow$ 'b $\Rightarrow$ 'b"
 fixes elem:: "'a  $\Rightarrow$ 'b $\Rightarrow$ bool"
 assumes
  empty_sem: "empty s = (sem s = {})"
                                    and
  nempty_sem: "nempty s = $\neg$ (empty s)" 
                                    and
  inter_sem: "sem (intersect s1 s2) =
               (sem s1) $\cap$ (sem s2)" and
  diff_sem: "sem (diff s1 s2) =
               (sem s1) - (sem s2)" and
  elem_sem: "elem a s $\equiv$  (a $\in$ sem s)"
\end{lstlisting}
\caption{The boolean algebra locale}
\label{fig-bool-algebra-locale}
\end{figure}

\paragraph{Implementation layer of the product operation.} We now present the algorithmic implementation of the product operation between an SFT and an SFA\footnote{For clarity of presentation, we show a simplified version of the Isabelle/HOL implementation while preserving the essential algorithmic structure.} based on Refined\_monadic framework.

Figure \ref{fig-compute-nft-product} illustrates the algorithm \texttt{productT\_impl}, which is implemented using the \emph{Refine\_Monadic} framework. Note that in the implementation, there are some refined operations:  \texttt{nfa\_states}, \texttt{nfa\_trans}, \texttt{nfa\_initial}, \texttt{nfa\_accepting},
and \texttt{nft\_tranfun}. These are corresponding to the states, transitions, initial states, accepting states, and output function mapping of the SFT.

The operation \texttt{prods\_imp}, shown in Figure \ref{fig-def-prods_imp}, computes the Cartesian product of two state sets. This function employs the \texttt{FOREACH} construct, a higher-order iteration operator analogous to OCaml's \texttt{Set.fold}. Specifically, given a set $S$, a function $f$ of type $'a \Rightarrow 'b \Rightarrow 'b$, and an initial accumulator $I$ of type $'b$, the expression $\texttt{FOREACH}~S~f~I$ systematically applies $f$ to each element in $S$, accumulating results in a principled manner.

\begin{figure}[hbt!]
	\begin{lstlisting}
definition productT_impl where
"productT_impl |$\mathcal{T}$| |$\mathcal{A}$| F fe = do {
    Q |$\leftarrow$| prods_imp (nft_states |$\mathcal{T}$|) 
    (nfa_states |$\mathcal{A}$|);
    (D1, D2) |$\leftarrow$| trans_comp_imp 
     (nft_tranfun |$\mathcal{T}$|) F fe (nft_trans |$\mathcal{T}$|) 
      (nfa_trans |$\mathcal{A}$|) (nfa_states |$\mathcal{A}$|);
    I |$\leftarrow$| prods_imp (nft_initial |$\mathcal{T}$|)
      (nfa_initial |$\mathcal{A}$|);
    F |$\leftarrow$| prods_imp (nft_accepting |$\mathcal{T}$|) 
      (nfa_accepting |$\mathcal{A}$|);
    RETURN (Q, D1, D2, I, F)
  }"
\end{lstlisting}
\caption{The computation of SFT product}
\label{fig-compute-nft-product}
\end{figure}

\begin{figure}[hbt!]
	\begin{lstlisting}
definition prods_imp where
"prods_imp Q1 Q2 =
   FOREACH {q. q |$\in$| Q1} (|$\lambda$| q Q. do {
    S |$\leftarrow$| FOREACH {q. q |$\in$| Q2}
      (|$\lambda$| q|$'$| Q|$'$|. RETURN ({(q,q|$'$|)} |$\cup$| Q|$'$|)) |$\emptyset$|;
    RETURN (Q |$\cup$| S)
   }) |$\emptyset$|"
\end{lstlisting}
\caption{The computation of Cartesian product of two state sets}
\label{fig-def-prods_imp}
\end{figure}

\begin{figure}[hbt!]
	\begin{lstlisting}
definition trans_comp_imp where
"trans_comp_imp M F fe T1 T2 Q =
  FOREACH {t. t |$\in$| T1}
   (|$\lambda$|(q,(|$\alpha$|,f),q|$'$|) (D1,D2). 
    (if (|$\alpha$|=None) then 
     (subtrans_comp_|$\varepsilon$| M q f q|$'$| F fe 
                              T2 D1 D2)
     else
     (subtrans_comp M q (the |$\alpha$|) f 
            q|$'$| F fe T2 D1 D2))) (|$\emptyset$|, |$\emptyset$|)"
\end{lstlisting}
\caption{The computation of \texttt{trans\_comp\_imp}}
\label{fig-def-prods-imp}
\end{figure}

A central algorithmic challenge in our implementation lies in the computation of transition sets $\texttt{D1}$ (corresponding to $\Delta_e$) and $\texttt{D2}$ (corresponding to $\Delta_e'$) in Figure \ref{fig-compute-nft-product}. As shown in Figure \ref{fig-def-prods-imp}, we implement this computation through the function \texttt{trans\_comp\_imp}, which computes the synchronization of transitions between the SFT and SFA. This function decomposes the synchronization process into two distinct cases, each handled by a specialized function:

\begin{enumerate}
  \item \texttt{subtrans\_comp\_$\varepsilon$}: Processes $\varepsilon$-transitions in the SFT, where transitions consume no input but may produce output
  \item \texttt{subtrans\_comp}: Processes standard transitions in the SFT, where both input consumption and output generation may occur
\end{enumerate}

\begin{figure}[hbt!]
\begin{lstlisting}
definition subtrans_comp where
"subtrans_comp M q |$\alpha$| f q|$'$| F fe T D1 D2 =
 FOREACH {t.t|$\in$|T} (|$\lambda$| (q1,|$\alpha'$|,q1|$'$|) (D1,D2).
  (if (nempty (intersect |$\alpha$| |$\alpha'$|)) then
    do {
      D1 |$\leftarrow$| 
       (if (F (M f) (intersect |$\alpha$| |$\alpha'$|)) 
          |$\neq$| None) then
         let |$\alpha_i$| = the (F (M f) 
              (intersect |$\alpha$| |$\alpha'$|)) in
         RETURN {((q,q1), |$\alpha_i$|, (q|$'$|,q1|$'$|))} 
                |$\cup$| D1
        else RETURN D1);
      D2 |$\leftarrow$| 
       (if fe (M f) (intersect |$\alpha$| |$\alpha'$|) 
        then 
         RETURN {((q,q1), (q|$'$|,q1|$'$|))} |$\cup$| D2 
        else RETURN D2);
      RETURN (D1, D2) }
   else (RETURN (D1, D2)))) (D1, D2)"
    \end{lstlisting}
    \caption{The computation of \texttt{subtrans\_comp}}
    \label{fig-def-subtrans_comp}
    \end{figure}

    We now present the implementation of \texttt{subtrans\_comp} in detail, as shown in Figure \ref{fig-def-subtrans_comp} (the implementation of \texttt{subtrans\_comp\_$\varepsilon$} follows analogous principles). For a transtion $(q, (\text{Some}~\alpha, f), q')$ in the SFT, this function traverses all transitions in the SFA, represented by the set \texttt{T}. For each transition $(q_1, \alpha', q_1')\in \texttt{T}$, the function performs two key operations when the intersection of input labels is non-empty ($\alpha \cap \alpha' \neq \emptyset$, verified using \texttt{nemptyIs}):

    \begin{enumerate}
      \item Computes non-$\varepsilon$-transitions ($\texttt{D1}$): When the output function applied to the intersection $\alpha \cap \alpha'$ yields a non-empty set, a new transition is added to $\texttt{D1}$ with the computed output label.
      \item Generates $\varepsilon$-transitions ($\texttt{D2}$): When there exists at least one input in the intersection $\alpha \cap \alpha'$ that produces an empty string (verified by checking if $\texttt{M}~f$ maps any element to \texttt{None}. The checking is implemented by \texttt{fe}), a corresponding $\varepsilon$-transition is added to $\texttt{D2}$.
    \end{enumerate}

  \begin{figure}[hbt!]
    \begin{lstlisting}
  lemma productT_imp_correct:
  assumes finite_TT: "|$\text{finite } (\Delta_t~\mathcal{T})$|"
      and finite_TA: "|$\text{finite } (\Delta~\mathcal{A})$|"
      and finite_Q: "|$\text{finite } (\mathcal{Q}~\mathcal{A})$|"
      and finite_TQ: "|$\text{finite } (\mathcal{Q}_t~\mathcal{T})$|"
      and finite_I: "|$\text{finite } (\mathcal{I}~\mathcal{A})$|"
      and finite_TI: "|$\text{finite } (\mathcal{I}_t~\mathcal{T})$|"
      and finite_F: "|$\text{finite } (\mathcal{F}~\mathcal{A})$|"
      and finite_TF: "|$\text{finite } (\mathcal{F}_t~\mathcal{T})$|"
  shows "|$\text{productT\_imp } \mathcal{T}~\mathcal{A}~F~\text{fe}\leq$|
            |$\text{SPEC } (\lambda A.~A = \text{productT } \mathcal{T}~\mathcal{A}~F)$|"
  \end{lstlisting}
  \caption{The refinement relation between \texttt{productT\_imp} and \texttt{productT}}
  \label{fig-def-productT_imp_correct}
  \end{figure}

  The correctness of the product computation is established through a refinement proof, demonstrating that \texttt{productT\_imp} (Figure \ref{fig-compute-nft-product}) correctly implements the abstract specification \texttt{productT} (Figure \ref{fig-def-FTProd}). This refinement relationship is formally specified in Figure \ref{fig-def-productT_imp_correct}, where we leverage the \emph{Refine\_Monadic} framework's data refinement.

  The refinement is expressed through the relation $\mathbf{C} \leq \texttt{SPEC}~\mathbf{A}$, which asserts that the concrete implementation $\mathbf{C}$ is an element of the abstract specification $\mathbf{A}$. More precisely,  the concrete implementation must produce an $\varepsilon$SFA that is structurally equivalent or isomorphic to the one produced by the abstract algorithm \texttt{productT}.

  To establish this equivalence, we must prove that the $\varepsilon$SFAs produced by \texttt{productT\_imp} and \texttt{productT} are isomorphic in all essential components: the set of states, transition relations, $\varepsilon$-transition relations, initial and accepting state sets.

The implementation of \texttt{productT\_imp} takes advantages of \emph{Refine\_Monadic} framework's interfaces for sets to store states and transition relations. 
The \emph{Refine\_Monadic} framework provides a way to automatically refine these interfaces to more efficient data structure, such as red-black trees or hashmaps. 
In our formalization, we refine the sets of storing states and transitions to red-black trees.

\subsection{Interval Implementation}

Finally, we present the implementation of the interval algebra which implements all operations defined in boolean algebra locale.
An interval is defined as a pair $(i, j)$ (represented as $[i-j]$ as well in the paper) representing the set $\{e \mid i \leq e \leq j\}$. To achieve greater expressiveness, our formalization extends this notion to interval lists of the form $[(i_1, j_1), \ldots, (i_n, j_n)]$, which denote the set $\bigcup_{1\leq k\leq n}\{e \mid i_k \leq e \leq j_k\}$. This generalization offers two key advantages: it enables more compact representation of transitions in SFAs and SFTs through merging, and it allows for efficient handling of interval operations without unnecessary splitting. For example, the set difference between intervals $(1, 5)$ and $(3, 4)$ can be directly represented as the interval list $[(1, 2), (5, 5)]$, which is also an interval.

Throughout the following discussion, we use the term "interval" to refer to interval lists. Our formalization provides all operations in the boolean algebra locale. For instance, we implement \texttt{intersect} for intervals, which computes the intersection of two intervals $i_1$ and $i_2$, yielding an interval $i$ such that $\texttt{sem}~i = \texttt{sem}~i_1 \cap \texttt{sem}~i_2$

To facilitate formal reasoning and optimize performance, in addition to the above locale operations, we introduce a canonical form for interval. An interval $[(i_1, j_1), $ $\ldots, (i_n, j_n)]$ is in canonical form if it satisfies two key properties:
\begin{enumerate}
  \item Each $(i_k, j_k)$ is well-formed: $i_k \leq j_k$ for all $k \in \{1,\ldots,n\}$
  \item Intervals are ordered and non-overlapping: $j_k < i_{k+1}$ for all $k \in \{1,\ldots,n-1\}$
\end{enumerate}

We prove that all interval operations preserve canonical form when applied to canonically-formed inputs. This invariant serves two purposes: it simplifies formal proofs by eliminating the need to reason about malformed or overlapping intervals, and it enables more efficient implementations of interval operations by reducing the number of cases to consider.

\section{Applications and Evaluation}
\label{sec:application}
We evaluate our formalization and implementation of the SFT product operation across two key application domains: string sanitization and string replacement in string solving. These case studies highlight the practical value of our certified SFT formalization in real-world scenarios.
\subsection{Sanitizer}

Sanitizers are crucial for preventing security vulnerabilities such as cross-site scripting (XSS) attacks by ensuring that user inputs are properly encoded before rendering in web browsers.

Listing \ref{lst-html-encode} presents a C\# code snippet implementing an HTML encoding function for Anti-XSS. This function iterates through each character in the input string and encodes characters that are unsafe for HTML rendering.

\begin{lstlisting}[language={[Sharp]C}, caption={C\# Code for AntiXSS.EncodeHtml version 2.0.}, label={lst-html-encode}, float=htbp]
static string EncodeHtml(string t)
{
  if (t == null) { return null; }
  if (t.Length == 0) {
    return string.Empty;
  }
  StringBuilder builder = new
    StringBuilder("", t.Length * 2);
  foreach (char c in t) {
    if ((((c > ''') && (c < '{')) |$\mid\mid$|
         ((c > '@') && (c < '['))) |$\mid\mid$|
         (((c == ' ') || ((c > '/') && (c < ':'))) |$\mid\mid$|
         (((c == '.') || (c == ',')) |$\mid\mid$|
         ((c == '-') || (c == '_'))))) {
          builder.Append(c);
      }
      else {
          builder.Append("&#" +
          ((int) c).ToString() + ";");
      }
  }
  return builder.ToString();
}
\end{lstlisting}

Figure \ref{fig:html-sanitizer-sft} illustrates the SFT representation of the HTML encoding sanitizer function.
 The set of safe characters is defined as $\texttt{safe\_chars} = [\texttt{' '}, \texttt{','}, \texttt{'.'}, \texttt{'-'}, \texttt{'\_'}] \cup [\texttt{'0'}-\texttt{'9'}] \cup [\texttt{'A'}-\texttt{'Z'}] \cup [\texttt{'a'}-\texttt{'z'}]$, which can be efficiently represented using the interval algebra we formalized. Unsafe characters are transformed using the encoding function $\texttt{encode}(c) = \texttt{``\&\#''} + \texttt{toString}(\texttt{ord}(c)) + \texttt{``;''}$. When the alphabet is set to Unicode $[\mathtt{0x0000}, \mathtt{0x10FFFF}]$, the complement $\neg\texttt{safe\_chars}$ represents all Unicode characters outside the safe set, which can also be efficiently represented using our formalized interval operations. \texttt{id}, \texttt{encode}, and $\lambda \_.\varepsilon$ are the identity, encode of unsafe characters, and empty word output functions, respectively.
\begin{figure}[htbp]
\centering
\begin{tikzpicture}[shorten >=1pt, node distance=2.5cm, on grid, auto]
  % States
  \node[state, initial, accepting] (q0) {$q_0$};
  \node[state] (q1) [right=3.5cm of q0] {$q_1$};
  \node[state] (q2) [below=of q0] {$q_2$};
  
  \path[->] 
    % Transition for safe characters
    (q0) edge[bend left=20] node[above,midway=1] {$\texttt{safe\_chars}, \texttt{id}$} (q1)
    % Transition for unsafe characters  
    (q0) edge[bend right=20] node[left,midway] {$\neg\texttt{safe\_chars}, \texttt{encode}$} (q2)
    % Epsilon transitions back to q0
    (q1) edge[bend left=30] node[below,midway] {$\varepsilon, \lambda \_.\varepsilon$} (q0)
    (q2) edge[bend right=20] node[right,midway] {$\varepsilon, \lambda \_.\varepsilon$} (q0);
\end{tikzpicture}
\caption{SFT representation of the HTML encoding sanitizer function.}
\label{fig:html-sanitizer-sft}
\end{figure}

We first evaluate the performance of our SFT product operation by computing the product of the HTML encoding sanitizer shown in Figure \ref{fig:html-sanitizer-sft} with SFAs constructed from randomly generated Unicode sequences. This product represents the sanitization result by the SFT for the input sequence. The experimental evaluation was conducted on a laptop equipped with an Apple M4 processor and 24 GB of memory, with a one-minute time limit imposed for each test.

\begin{figure}
\centering
\includegraphics[scale=0.45]{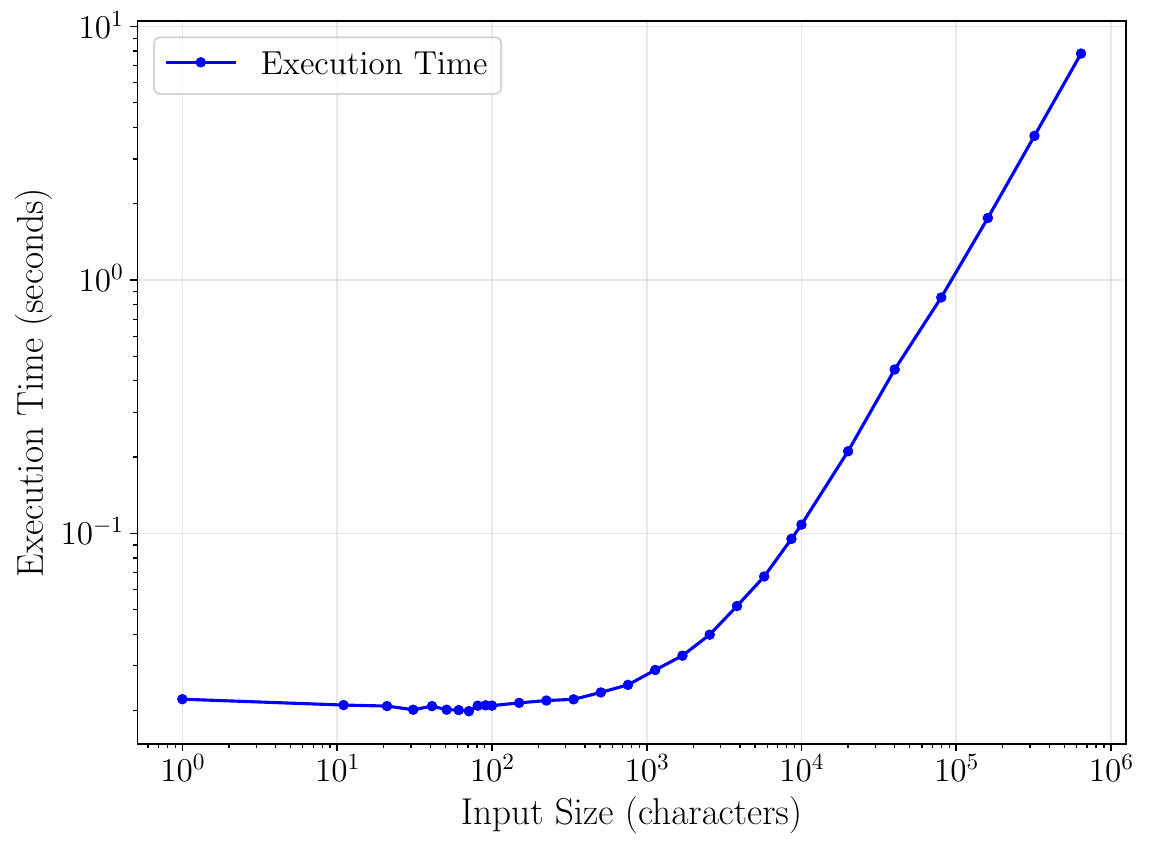}
\caption{Experimental results for the sanitizer product}
\label{fig:benchmark-results}
\end{figure}

Figure \ref{fig:benchmark-results} demonstrates the scalability characteristics of our SFT product implementation. The computation requires approximately 1 second for 100,000-character inputs and 10 seconds for 1,000,000-character inputs. These performance metrics confirm that our certified implementation maintains computational efficiency suitable for practical string processing applications at scale.

Beyond performance evaluation, our SFT formalization also supports the formal verification of string sanitizer's safety property. A sanitizer transforms user inputs into safe outputs, ensuring that the result contains no unsafe characters. However, sanitizers—like any other programs—are error-prone. Manual reviews or testing alone are often insufficient to guarantee their correctness.

Based on our formalization, we can verify sanitizers by modeling the sanitizer as an SFT and the user input as an SFA. We can also model potential attacks as SFAs that accept unsafe strings. Formally speaking, Let $\mathcal{T}$ be an SFT, $\mathcal{I}$ is the possible user inputs modeled by an SFA, and $\mathcal{A}$ is the attack SFA that accepts unsafe strings. The santizer correctness checking against the attack is to check whether the intersection of the product $\mathcal{T} \times \mathcal{I}$ and the attack SFA $\mathcal{A}$ is empty, i.e., $(\mathcal{T} \times \mathcal{I}) \cap \mathcal{A} = \emptyset$.

Moreover, santizers are usually used in combination with other sanitizers to ensure comprehensive safety. For example, the HTML encoding sanitizer can be combined with a string trimming operation to remove leading and trailing whitespace characters. For this kind of sanitizer composition, it can be modeled by our SFT product as well. Let $\mathcal{T}_1$ and $\mathcal{T}_2$ correspond to two sanitizers, and $\mathcal{T}_1$ is applied before $\mathcal{T}_2$. What we need to check is $(\mathcal{T}_2 \times (\mathcal{T}_1 \times \mathcal{I})) \cap \mathcal{A} = \emptyset$.

\begin{table}[h]
  \centering
  \small
  \setlength{\tabcolsep}{4pt} % Reduced column gap
  \begin{tabular}{p{2.2cm}ll} % Increased width for automatic line break
    \toprule
    \textbf{Attack Model} & \textbf{Verfication Result} & \textbf{Size} \\
    \midrule
    attack\_html           & GA HtmlEscape: \textbf{Unsafe} & 1/6\\
                           & Trim + GA HtmlEscape: \textbf{Safe} & 4/12 \\
                           & escapeString: \textbf{Unsafe} & 1/11 \\
                           & Trim + escapeString: \textbf{Safe} & 4/17 \\
                           & Trim + OWASP HTMLEncode: \textbf{Safe} & 4/18 \\
    \midrule
    attack\_javascript     & GA HtmlEscape + GA PreEscape + & 3/20 \\
                           &  gaSnippetesc: \textbf{safe} & \\
    \midrule
    attack\_json           & Trim + GA JSONESc : \textbf{Safe} & 4/16\\
    \midrule
    attack\_xml            & Trim + GA XMLEsc: \textbf{safe} & 4/16\\
    \midrule
    attack\_ccs            & Trim + GA CleanseCSS : \textbf{Safe} & 25/46\\
    \bottomrule
  \end{tabular}
  \caption{Sanitizer verfication result}
  \label{tab:sanitizer_attack}
\end{table}

Table \ref{tab:sanitizer_attack} summarizes the verification results for various sanitizers against different attack models. The first column lists the attack models, the second column indicates whether the sanitizer is classified as safe or unsafe with respect to the attack, and the third column reports the size of the transducers used in the verification, given as [number of states/number of transitions]. For composed SFTs, the size reflects the total number of states and transitions across all SFTs.

We evaluate a range of attack models for HTML, JavaScript, JSON, XML, and CSS. Listing \ref{lst-css-attack} shows the CSS attack model, which is used to verify the corresponding sanitizer. The attack model includes comments, declaration breakers, and known dangerous tokens such as \textbf{expression(}, \textbf{@import}, \textbf{behavior:}, \textbf{-moz-binding}, and \textbf{url(} (unless the URL is separately validated).
These are unsafe and undesired strings in CSS contexts.
Sanitizers include GA HtmlEscape, GA JSONEscape, GA XMLEscape, escapeString, and CleanseCSS, where "GA" means from Google Analytics. Sanitizers are often composed with preprocessing steps such as Trim, which removes leading and trailing whitespace characters. As shown in the table, GAHtmlEscape and escapeString are unsafe with respect to the HTML attack model without preprocessing, but become safe when combined with Trim. The \textbf{Size} column demonstrates the succinctness of SFTs in modeling real-world transducers. We did not list the execution time of verification because all of them are quite efficent with less than 0.02 second to finish the verificaiton.

\begin{lstlisting}[language={}, caption={CSS attack model for sanitizer verification.}, label={lst-css-attack}, float=htbp]
|$\Sigma$|* ( / \* [\s\S]*? \*/
  |$\mid$| ; |$\mid$| \{ |$\mid$| \}
  |$\mid$| [eE][xX][pP][rR][eE][sS][sS][iI]
    [oO][nN]\s*\(
  |$\mid$| @\s*[iI][mM][pP][oO][rR][tT]
  |$\mid$| [bB][eE][hH][aA][vV][iI][oO][rR]\s*:
  |$\mid$| -\s*[mM][oO][zZ]-\s*[bB][iI][nN]
    [dD][iI][nN][gG]
  |$\mid$| (?!allowUrl) [uU][rR][lL]\s*\(
 ) |$\Sigma$|*
\end{lstlisting}

%\subsection{Modeling the Replacement Operation}

\subsection{Applications to String Replacement}

The string replacement, denoted as
$\texttt{replace}(\texttt{str}, \texttt{pattern}, \newline\texttt{replacement}, \texttt{flag})$, is a fundamental string transformation accepting four parameters:
(1) $\texttt{str}$: The input string to be transformed.
(2) $\texttt{pattern}$: A regular expression defining the matching pattern.
(3) $\texttt{replacement}$: The string that replaces matched substrings.
(4) $\texttt{flag}=s/g$: The replacement mode flag, where $s$ denotes single first occurrence replacement and $g$ denotes all occurrences replacement operation.

% The operation's semantics is to replace the first occurrence (or all occurrences) of any substring $s'$ in \texttt{str} that matches $\texttt{pattern}$ with \texttt{replacement}.

To illustrate how SFTs model this operation, consider $\texttt{replace}(s, \texttt{/[0-9]+/},~$\texttt{NUM}$,~\texttt{flag})$. We construct the corresponding SFT by the following three steps.

\begin{enumerate}
  \item Construct an SFA recognizing the pattern $\texttt{/[0-9]+/}$ and transform it into an SFT by augmenting each transition with output function "$\lambda x.~\texttt{None}$" that produces the empty string.
  \item Construct an SFA accepting the replacement string \texttt{NUM} and convert it to an SFT by
  replacing each transtion in the SFA $(q, c, q')$ with
  $(q, (\texttt{None},~\lambda \_.~c ), q')$.
  \item Compose the two SFTs by connecting the accepting state of the pattern-matching SFT to the initial state of the replacement-generating SFT via $\varepsilon$-transitions.
\end{enumerate}

%\noindent\emph{Step 1.}
%Figure \ref{fig-snfa-pattern} illustrates the construction of the pattern-matching component. The left side shows the SFA that recognizes the regular expression $\texttt{/[0-9]+/}$, while the right side presents its transformation into an SFT. This transformation is achieved by augmenting each transition with the output function $\texttt{f} = \lambda x.~\texttt{None}$, which consistently produces the empty string, effectively "consuming" the matched digits without generating output.

Figure \ref{fig-rearranged-automata} presents the complete SFT construction, obtained by composing the pattern-matching SFT generated from \texttt{/[0-9]+/}  with the replacement-generating SFT generated from \texttt{NUM}.

\begin{figure}[h] \centering
  \begin{tikzpicture}[shorten >=1pt, node distance=2cm, auto]
    % Second automaton from the first figure
    \node[state, initial] (q0') {$q_0'$}; 
    \node[state] (q1') [right=of q0'] {$q_1'$}; 
  
     \path[->] 
     (q0') edge [bend left] node {\texttt{[0-9]}, $\lambda~x.~\texttt{None}$} (q1')
     (q1') edge [loop right] node {\texttt{[0-9]}, $\lambda~x.~\texttt{None}$} (q1');
  
    % Second automaton from the second figure
    \node[state] (q0'') [below=1cm of q1'] {$q_0''$}; 
    \node[state] (q1'') [right=of q0''] {$q_1''$}; 
    \node[state, accepting] (q3'') [below=1cm of q0''] {$q_3''$}; 
    \node[state] (q2'') [right=of q3''] {$q_2''$}; 
  
    \path[->] 
    (q0'') edge node {\texttt{None}, $\lambda\_.\texttt{N}$} (q1'')
    (q1'') edge [bend left=5] node[left] {\texttt{None}, $\lambda\_.\texttt{U}$} (q2'')
    (q2'') edge node {\texttt{None}, $\lambda\_.\texttt{M}$} (q3'')
    (q1') edge [bend right=30] node {\texttt{None}, $\lambda x.~\texttt{None}$} (q0'');
  \end{tikzpicture}
  \caption{The SFT for the replacement operation $\texttt{replace}(s,\texttt{/[0-9]+/}, \texttt{NUM})$}
  \label{fig-rearranged-automata}
  \end{figure}

  We can extend the above construction to model both replace-first and replace-all for XSS attack verification. In this paper, we only illustrate the construction of replace-all in Figure \ref{fig:replace-all}. The SFT $\mathcal{T}_2$ combines pattern matching and replacement using the construction just described. The SFT $\mathcal{T}_1$ matches strings that do not contain any substrings matching the pattern. Two $\varepsilon$-transitions connect these components: one from $\mathcal{T}_1$'s accepting state to $\mathcal{T}_2$'s initial state, and another from $\mathcal{T}_2$'s accepting state back to $\mathcal{T}_1$'s initial state. This cyclic structure enables the replace-all operation by repeatedly applying pattern replacement until no more matches exist in the input string. Replace-first modeling has no transtion from $\mathcal{T}_2$'s accepting state back to $\mathcal{T}_1$'s initial state.

\begin{figure}[htbp]
\centering
\begin{tikzpicture}[node distance=2cm, state/.style={circle, draw, minimum size=0.8cm}]
  % First SFT rectangle
  \node[draw, rectangle, minimum width=5.2cm, minimum height=2.1cm] (sft1-box) {};
  \node[above=0.1cm of sft1-box.north] {\textbf{SFT} $\mathcal{T}_1$ (Non-matching)};
  
  % States inside first SFT
  \node[state, initial] (p0) at ([xshift=-0.8cm]sft1-box.center) {$p_0$};
  \node[state, accepting] (p1) at ([xshift=1.8cm]sft1-box.center) {$p_1$};
  \path[->] (p0) edge[above] node {\shortstack{$\neg\texttt{pattern},$ \\ $\lambda x.x$}} (p1);
  
  % Second SFT rectangle
  \node[draw, rectangle, minimum width=5.2cm, minimum height=2.1cm, below=1.5cm of sft1-box] (sft2-box) {};
  \node[below=0.1cm of sft2-box.south] {\textbf{SFT} $\mathcal{T}_2$ (Matching)};
  
  % States inside second SFT
  \node[state, initial] (s0) at ([xshift=-0.8cm]sft2-box.center) {$s_0$};
  \node[state, accepting] (s1) at ([xshift=2cm]sft2-box.center) {$s_1$};
  \path[->] (s0) edge[above] node {\shortstack{$\texttt{pattern},$ \\ $\texttt{replacement}$}} (s1);
  
  % Epsilon transition from accepting state of first SFT to initial state of second SFT
  \path[->] (p1) edge[bend right=30] node[left] {\shortstack{$\varepsilon$, $\lambda\varepsilon$}} (s0);
  
  % Epsilon transition from accepting state of second SFT to initial state of first SFT
  \path[->] (s1) edge[bend right=30] node[right] {\shortstack{$\varepsilon$, $\lambda\varepsilon$}} (p0);
  
\end{tikzpicture}
\caption{The modeling of replace all operation.}
\label{fig:replace-all}
\end{figure}

We evaluate our SFT-based replacement operation by extending CertiStr \cite{cpp/KanLRS22}, a certified string solver that previously supported only basic string operations: concatenation and regular constraints. The original CertiStr was incapable of verifying any XSS attack due to the lack of support for transducers and replacement operations.
CertiStr employs a forward-propagation approach to solve string constraints, which requires computing the forward image of string operations. Our SFT product operation provides an exact forward image computation for replacement operations, enabling seamless integration of replacement functionality into CertiStr's solving framework.

By incorporating our SFT-based modeling of replacement, we develop CertiStrR—an
extended string solver that supports two types of string replacement: (1)
replacing the first occurrence of a substring matching a given pattern, and (2)
replacing all occurrences of substrings matching a given pattern, i.e., either
a constant string or a regular expression.

%While the core solving algorithm maintains the certification guarantees of CertiStrR, the frontend components are implemented using established non-certified  OCaml libraries: dolmen \cite{dolmen} for SMT-LIB parsing and ocaml-re-nfa \cite{ocaml-re-nfa} for regular expression to NFA conversion.

% Listing \ref{lst-smtlib-code} demonstrates the regular expression variant through an example that replaces numeric sequences with the string "\texttt{NUM}". The constraints are satisfiable because the input string \texttt{a} = "\texttt{2024,2025}" contains a substring "\texttt{2025}" that matches the regular expression \texttt{re.+ (re.range "0" "9")} (equivalent to \texttt{/[0-9]+/}), and replacing this match with "\texttt{NUM}" yields the expected output string \texttt{b} = "\texttt{2024,NUM}".
% %
%
%
% \begin{lstlisting}[language=SMTLIB, caption={Example SMT-LIB Code}, label={lst-smtlib-code}, float=htbp]
%   (set-logic QF_S)
%   (declare-fun a () String)
%   (declare-fun b () String)
%   (assert (= a "2024,2025"))
%   (assert (= b "2024,NUM"))
%   (assert (= b (str.replace_re a 
%       (re.+ (re.range "0" "9")) "NUM")))
%   (check-sat)
%   \end{lstlisting}

We evaluate {CertiStrR} using benchmarks from SMT-LIB 2025~\cite{smtlib_benchmarks}, focusing on the \textsf{QF\_S} and \textsf{QF\_SLIA} logic fragments. The benchmarks are divided into two categories: (1) replace-first and (2) replace-all. In addition to replacement, the tests also include string concatenation and regular constraints. Due to the limitation of CertiStrR's front-end, some constraints were rewritten or removed from the tests.

% replace_re
% Average real time: 0.37 seconds
% Total number of real entries: 98
% Average user time: 0.35 seconds
% Total number of user entries: 98
% Average sys time: 0.01 seconds
% Total number of sys entries: 98
% Total number of SAT entries: 61
% Total number of UNSAT entries: 4
% Total number of inconclusive entries: 30

% replace_str
% Average real time: 0.27 seconds
% Total number of real entries: 323
% Average user time: 0.25 seconds
% Total number of user entries: 323
% Average sys time: 0.01 seconds
% Total number of sys entries: 323
% Total number of SAT entries: 315
% Total number of UNSAT entries: 173
% Total number of inconclusive entries: 8

%\begin{table}[h]
%  \centering
%  \small
%  \begin{tabular}{lcccc}
%      \toprule
%      & \textbf{SAT/UNSAT/Inc} & \textbf{Time (s)} & \textbf{Tests} %\\
%      \midrule
%      \texttt{replace singleton} & 142/173/8 & 0.27 & 323\\
%      \texttt{replace all} & 84/4/10 & 0.36 & 98\\
%      \bottomrule
%  \end{tabular}
%  \caption{Experimental results}
%  \label{tab:string_operations}
%\end{table}

\begin{table*}[h]
\centering
\small
\setlength{\tabcolsep}{4pt}
\renewcommand{\arraystretch}{1.1}
\begin{tabular}{llccccccc}
  \toprule
  \textbf{Subgroup} & \textbf{Test} & \textbf{Count} & \textbf{SAT} & \textbf{UNSAT} & \textbf{Inconclusive} & \textbf{CertiStrR (s)} & \textbf{Ostrich (s/timeout)} & \textbf{CVC5 (s/timeout)} \\
  \midrule
  \multirow{2}{*}{Replace-First} & Replace RE  & 94  & 77  & 16  & 1 & 0.79 & 166.27/27 & 2.67/20 \\
   & Replace Str & 323 & 103 & 211 & 9 & 1.94 & 7.10/20 & 0.81/1 \\
     \midrule
  \multirow{4}{*}{Replace-All} & PCP & 1000 & 751 & 249 & 0 & 0.14 & 3.412/0 & unknown \\
   & RNA SAT   & 500  & 500 & 0   & 0 & 2.81 & 55.61/0 & unknown \\
   & RNA UNSAT & 500  & 0   & 500 & 0 & 3.14 & 64.185/0 & unknown \\
   & WebApp      & 255  & 40  & 215 & 0 & 0.79 & 23.67/17 & 8.379/1 \\
  \bottomrule
\end{tabular}
\caption{Experimental results}
\label{tab:string_operations}
\end{table*}

The experimental evaluation was conducted on a laptop with an Apple M4 processor and 24 GB of memory, with a \textbf{one-hour} time limit per test. Table~\ref{tab:string_operations} summarizes the results. The results from CertiStrR were classified into three categories: SAT (satisfiable), UNSAT (unsatisfiable), and Inconclusive. An "Inconclusive" result indicates that the solver cannot determine satisfiability due to the inherent incompleteness of the string solver's algorithm, not timeout (all tests completed within one hour). This occurs when the solver cannot determine whether string variables need to be assigned different values that satisfy all constraints. 

We also compare our approach with two state-of-the-art string solvers: Ostrich \cite{pacmpl/ChenFHHHKLRW22} and CVC5 \cite{cvc5}.
The experimental results demonstrate that string replacement operations, particularly replace-all operations, pose significant challenges for existing state-of-the-art solvers.
The "unknown" label in the table indicates the tests where CVC5 cannot determine satisfiability. 
Notably, PCP and RNA benchmarks cannot be solved by most state-of-the-art string
solvers, including CVC5 \cite{cvc5}, yet CertiStrR solves
them efficiently. Among existing solvers, only OSTRICH \cite{pacmpl/ChenFHHHKLRW22} can handle these benchmarks.
CertiStrR outperforms both Ostrich and CVC5 across most benchmark subgroups, with the only exception of the "Replace Str" category where CVC5 achieves better performance (though it experiences one timeout). CertiStrR demonstrates particularly strong performance in  "Replace All" category.
These results demonstrate the effectiveness and efficiency of our SFT formalization and implementation.

\paragraph{Transducers vs. Replacement Operations.}
Although both transducers and replacement operations can express string
transformations for security-related verification, transducers are strictly more
expressive because they can capture dependencies between the input pattern and
the replacement string. In contrast, all replacement operations in SMT-LIB treat
the pattern and the replacement string as independent. For example, in the SFT
in Figure~\ref{fig:html-sanitizer-sft}, a transition may be defined over
an interval such as $[\texttt{a-z}]$, and the output function computes its
output directly based on the input character. Replacement operations, however,
cannot express such dependencies since the replacement string is not a function.

Consider a string transformation that replace each character in $[\texttt{a-z}]$ with its ASCII code. Modeling this transformation using only replacement operations requires spliting the interval $[\texttt{a-z}]$ into individual characters, resulting in 26 separate \texttt{replace-all} operations—one for each letter. For example:
\begin{verbatim}
  replace(replace(..., b, 0x62, g), a, 0x61, g)
\end{verbatim}

When the interval contains a large set of characters—as is common in web applications that support the full Unicode range of 
\texttt{0x10FFFF} code points—this nested structure becomes impractical, leading to significant inefficiency and complexity in the replacement-based approach.

\subsection{Effort of Certified Development}

%This subsection discusses the development effort required for our certified SFT formalization. 
While SFTs constitute the central contribution of this paper, they are not sufficient on their own to support the applications we target. A more comprehensive symbolic framework for automata libraries is necessary. Beyond the existing automata library supported by CertiStr \cite{cpp/KanLRS22}, which already includes operations such as SFA product and SFA concatenation, we needed to develop additional \emph{certified} SFA operations from scratch. These include $\varepsilon$SFAs, translation from $\varepsilon$SFAs to SFAs, translation from SFAs to Symbolic Deterministic Finite Automata (SDFA), and negation of SFAs. Table~\ref{tab:abstract_impl} summarizes the certified development effort that was carried out specifically for this work.

The abstract-level development encompasses all formalizations presented in Section~\ref{sec:abstract-layer}, including SFTs and the additional SFA operations used in this paper's application. The implementation-level development covers all formalizations described in Section~\ref{sec_alg_refinement}, including the refinement of the product operation and other SFA operations used in this paper's application. The final row of Table~\ref{tab:abstract_impl} corresponds to the effort devoted to interval formalization. The most challenging component proved to be the correctness proof of the product operation at the abstract level (Figure~\ref{fig-def-product-correct}).
%
%Compared to CertiStr \cite{cpp/KanLRS22}, the interval formalization in CertiStrR is substantially more complex, primarily because intervals were extended from single pairs to lists.

\begin{table}[h]
  \centering
  \small
  \begin{tabular}{lccc}
      \toprule
      & \textbf{Definitions} & \textbf{Lemmas} & \textbf{Proofs (LOC)} \\
      \midrule
      Abstract & 17 & 21 & 3274 \\
      Implementation & 51 & 43 & 2700 \\
      Interval & 15 & 29 & 1500 \\
      \bottomrule
  \end{tabular}
  \caption{Overview of the effort of certified development}
  \label{tab:abstract_impl}
\end{table}

\section{Related Work}
\label{sec:related-work}

\emph{Symbolic Automata and Transducers.} Symbolic Automata and Transducers \cite{cav/DAntoniV17,VeanesHLMB12Transducer, popl/DAntoniV14, entcs/DAntoniKW18, sofsem/TammV18} represent a significant advancement in automata theory, offering improved efficiency in operations and enhanced expressiveness through algebraic theories that support infinite alphabets. This symbolic framework has been progressively extended to accommodate more complex structures: symbolic tree transducers \cite{ershov/VeanesB11} handle hierarchical data structures, while symbolic pushdown automata \cite{cav/DAntoniA14} manage nested word structures. 
%Recent developments in 2024 have further expanded the scope of symbolic techniques to include B\"{u}chi automata and omega-regular languages \cite{pacmpl/VeanesBEZ25}, enabling verification of infinite-state systems and analysis of non-terminating computations.

\emph{Applications of Symbolic Transducers.} The efficiency, scalability, and
expressive power of symbolic automata and transducers have led to their
widespread adoption in numerous applications, including: constraint solving for program analysis \cite{lpar/VeanesBM10}, security-critical sanitizer analysis for web applications \cite{uss/HooimeijerLMSV11}, runtime verification of system behaviors \cite{osdi/YaseenABCL20}, automated program inversion for software 
transformation \cite{pldi/HuD17}, and string solving
\cite{pacmpl/ChenFHHHKLRW22,CHL+19}. Each application leverages the symbolic 
approach's ability to handle complex
patterns and large alphabets efficiently.

\emph{Formalization of Symbolic Automata and Transducers.} While classical automata theories have been extensively formalized in interactive theorem provers \cite{Tuerk-NFA, Lammich2014TheCA, Peter14, cpp/DoczkalKS13}, with some work on transducer formalization \cite{afp/LochmannFSTS21}, the symbolic variants of automata and transducers remain largely unexplored in formal verification. 
%To our knowledge, CertiStr \cite{cpp/KanLRS22} represents the only existing work on symbolic automata formalization in a proof assistant. Our work advances this frontier by extending the formal treatment to both SFTs and $\varepsilon$SFAs.

\emph{String Solving.} A significant application of our certified transducer
framework is in string constraint solving, a field that has seen intensive
research development over the past decade. While our work provides formal
verification guarantees, there exists a rich ecosystem of non-certified string
solvers, which has in the last few years led to a standardized SMT-LIB 2.6/2.7 
unicode string theory and annual competition. SMT-LIB compatible string 
solvers include Z3 \cite{Z3}, Z3-alpha \cite{z3alpha}, Z3-Noodler 
\cite{noodler,noodler-int,noodler-len, noodler-tool}, Z3str3RE \cite{z3str3re},
cvc5 \cite{cvc5}, Z3str4 \cite{z3str4}, Trau \cite{Z3-trau,trau-tool}, and 
OSTRICH \cite{ostrich,CHL+19,pacmpl/ChenFHHHKLRW22}. 
Recent SMT competition over string theory has revealed
superiority of automata-based methods (based on restrictions of symbolic
automata). 
We envision that our work will benefit the string solving community by 
providing a formal foundation for string solvers development.
%: Kaluza \cite{Berkeley-JavaScript} specializes in JavaScript analysis, CVC5 \cite{cvc5}, Z3-str3 \cite{Z3-str3} builds on the Z3 framework, S3P \cite{DBLP:conf/ccs/TrinhCJ14}, Ostrich \cite{pacmpl/ChenFHHHKLRW22}, and SLOTH \cite{HJLRV18}.
%As more and more 
%Numerous bugs have been uncovered in existing string solvers \cite{DBLP:conf/cav/BlotskyMBZKG18} and SMT solvers \cite{Mansur20} 

\emph{Certified SMT Solvers.} Certification efforts 
for SMT (other than string theories) also exist. For example, the work by Shi et al. \cite{DBLP:conf/cav/ShiFLTWY20}, who developed a certified SMT solver for quantifier-free bit-vector theory, demonstrating the broader applicability of interactive theorem proving in certified SMT solver development. The work by Mansur et al. \cite{verified-verifying} define a formal semantics for SMT-LIB string
to check compatibility between solvers and SMTLIB standard.
Finally, there are also recent works on certified proof checkers for automated 
reasoning and in particular SMT (cf. \cite{cacm-certified}).

\section{Conclusion}
\label{sec:conclusion}

We have presented the first formalization of symbolic finite transducers (SFTs)
and their most crucial algorithms in Isabelle/HOL. Our framework offers 
flexible interfaces that enable diverse applications, featuring support for 
$\varepsilon$-transitions on both inputs and outputs, as well as extensibility to various Boolean algebras through a refinement framework.
To demonstrate the practical utility of our formalization, we applied formalized
SFTs for analyzing sanitization for web applications, as well as certified 
string solving for complex constraints involving equality, concatenation,
regular constraints, and replaceall, for which we confirm the efficiency
and effectiveness of our approach on SMT-LIB 2025 benchmarks
\cite{smtlib_benchmarks}.
%We envision that our work will benefit the string solving community by 
%providing a formal foundation for string solvers development.
%from SMT-LIB 2025 \cite{smtlib_benchmarks}, and the experimental results confirm both the efficiency and effectiveness of our approach.
%applied SFTs to model string sanitization functions commonly used in web applications and showed how to verify the correctness of these sanitizers. Furthermore, we developed CertiStrR, an extension of CertiStr \cite{cpp/KanLRS22}, which adds support for string replacement operations. 

\bibliography{literature}

\begin{thebibliography}{10}

\bibitem{trau-tool}
Parosh~Aziz Abdulla, Mohamed~Faouzi Atig, Yu-Fang Chen, Bui~Phi Diep, Lukas Holik, Ahmed Rezine, and Philipp Ruemmer.
\newblock Trau: Smt solver for string constraints.
\newblock {\em 2018 Formal Methods in Computer Aided Design (FMCAD)}, pages 1--5, 2018.
\newblock URL: \url{https://api.semanticscholar.org/CorpusID:53962814}.

\bibitem{DBLP:conf/fmcad/BackesBCDGLRTV18}
John Backes, Pauline Bolignano, Byron Cook, Catherine Dodge, Andrew Gacek, Kasper~S{\o}e Luckow, Neha Rungta, Oksana Tkachuk, and Carsten Varming.
\newblock Semantic-based automated reasoning for {AWS} access policies using {SMT}.
\newblock In Nikolaj Bj{\o}rner and Arie Gurfinkel, editors, {\em 2018 Formal Methods in Computer Aided Design, {FMCAD} 2018, Austin, TX, USA, October 30 - November 2, 2018}, pages 1--9. {IEEE}, 2018.
\newblock \href {https://doi.org/10.23919/FMCAD.2018.8602994} {\path{doi:10.23919/FMCAD.2018.8602994}}.

\bibitem{cacm-certified}
Haniel Barbosa, Clark~W. Barrett, Byron Cook, Bruno Dutertre, Gereon Kremer, Hanna Lachnitt, Aina Niemetz, Andres N{\"{o}}tzli, Alex Ozdemir, Mathias Preiner, Andrew Reynolds, Cesare Tinelli, and Yoni Zohar.
\newblock Generating and exploiting automated reasoning proof certificates.
\newblock {\em Commun. {ACM}}, 66(10):86--95, 2023.
\newblock \href {https://doi.org/10.1145/3587692} {\path{doi:10.1145/3587692}}.

\bibitem{z3str3re}
Murphy Berzish, Mitja Kulczynski, Federico Mora, Florin Manea, Joel~D. Day, Dirk Nowotka, and Vijay Ganesh.
\newblock An {SMT} solver for regular expressions and linear arithmetic over string length.
\newblock In Alexandra Silva and K.~Rustan~M. Leino, editors, {\em Computer Aided Verification - 33rd International Conference, {CAV} 2021, Virtual Event, July 20-23, 2021, Proceedings, Part {II}}, volume 12760 of {\em Lecture Notes in Computer Science}, pages 289--312. Springer, 2021.
\newblock \href {https://doi.org/10.1007/978-3-030-81688-9\_14} {\path{doi:10.1007/978-3-030-81688-9\_14}}.

\bibitem{noodler}
Franti{\v{s}}ek Blahoudek, Yu-Fang Chen, David Chocholat{\'y}, Vojt{\v{e}}ch Havlena, Luk{\'a}{\v{s}} Hol{\'i}k, Ond{\v{r}}ej Leng{\'a}l, and Juraj S{\'i}{\v{c}}.
\newblock Word equations in synergy with regular constraints.
\newblock In Marsha Chechik, Joost-Pieter Katoen, and Martin Leucker, editors, {\em Formal Methods}, pages 403--423, Cham, 2023. Springer International Publishing.

\bibitem{DBLP:conf/cav/BlotskyMBZKG18}
Dmitry Blotsky, Federico Mora, Murphy Berzish, Yunhui Zheng, Ifaz Kabir, and Vijay Ganesh.
\newblock Stringfuzz: {A} fuzzer for string solvers.
\newblock In Hana Chockler and Georg Weissenbacher, editors, {\em Computer Aided Verification - 30th International Conference, {CAV} 2018, Held as Part of the Federated Logic Conference, FloC 2018, Oxford, UK, July 14-17, 2018, Proceedings, Part {II}}, volume 10982 of {\em Lecture Notes in Computer Science}, pages 45--51. Springer, 2018.
\newblock \href {https://doi.org/10.1007/978-3-319-96142-2\_6} {\path{doi:10.1007/978-3-319-96142-2\_6}}.

\bibitem{Peter14}
Julian Brunner.
\newblock {Transition Systems and Automata Isabelle Library}.
\newblock Arch. Formal Proofs, 2017.
\newblock \url{https://www.isa-afp.org/entries/Transition_Systems_and_Automata.html}.

\bibitem{BM20}
Alexandra Bugariu and Peter M{\"{u}}ller.
\newblock Automatically testing string solvers.
\newblock In Gregg Rothermel and Doo{-}Hwan Bae, editors, {\em {ICSE} '20: 42nd International Conference on Software Engineering, Seoul, South Korea, 27 June - 19 July, 2020}, pages 1459--1470. {ACM}, 2020.
\newblock \href {https://doi.org/10.1145/3377811.3380398} {\path{doi:10.1145/3377811.3380398}}.

\bibitem{Z3-trau}
Diep Bui and contributors.
\newblock Z3-trau.
\newblock \url{https://github.com/diepbp/z3-trau}, 2019.

\bibitem{pacmpl/ChenFHHHKLRW22}
Taolue Chen, Alejandro Flores{-}Lamas, Matthew Hague, Zhilei Han, Denghang Hu, Shuanglong Kan, Anthony~W. Lin, Philipp R{\"{u}}mmer, and Zhilin Wu.
\newblock Solving string constraints with regex-dependent functions through transducers with priorities and variables.
\newblock {\em Proc. {ACM} Program. Lang.}, 6({POPL}):1--31, 2022.
\newblock \href {https://doi.org/10.1145/3498707} {\path{doi:10.1145/3498707}}.

\bibitem{CHL+19}
Taolue Chen, Matthew Hague, Anthony~W. Lin, Philipp R{\"{u}}mmer, and Zhilin Wu.
\newblock Decision procedures for path feasibility of string-manipulating programs with complex operations.
\newblock {\em Proc. {ACM} Program. Lang.}, 3({POPL}):49:1--49:30, 2019.
\newblock \href {https://doi.org/10.1145/3290362} {\path{doi:10.1145/3290362}}.

\bibitem{noodler-len}
Yu-Fang Chen, David Chocholaty, Vojtech Havlena, Lukas Holik, Ondrej Lengal, and Juraj Sic.
\newblock Solving string constraints with lengths by stabilization.
\newblock {\em Proc. ACM Program. Lang.}, 7(OOPSLA2), oct 2023.
\newblock \href {https://doi.org/10.1145/3622872} {\path{doi:10.1145/3622872}}.

\bibitem{noodler-tool}
Yu-Fang Chen, David Chocholat{\'y}, Vojt{\v{e}}ch Havlena, Luk{\'a}{\v{s}} Hol{\'i}k, Ond{\v{r}}ej Leng{\'a}l, and Juraj S{\'i}{\v{c}}.
\newblock Z3-noodler: An automata-based string solver.
\newblock In Bernd Finkbeiner and Laura Kov{\'a}cs, editors, {\em Tools and Algorithms for the Construction and Analysis of Systems}, pages 24--33, Cham, 2024. Springer Nature Switzerland.

\bibitem{cav/DAntoniA14}
Loris D'Antoni and Rajeev Alur.
\newblock Symbolic visibly pushdown automata.
\newblock In Armin Biere and Roderick Bloem, editors, {\em Computer Aided Verification - 26th International Conference, {CAV} 2014, Held as Part of the Vienna Summer of Logic, {VSL} 2014, Vienna, Austria, July 18-22, 2014. Proceedings}, volume 8559 of {\em Lecture Notes in Computer Science}, pages 209--225. Springer, 2014.
\newblock \href {https://doi.org/10.1007/978-3-319-08867-9\_14} {\path{doi:10.1007/978-3-319-08867-9\_14}}.

\bibitem{entcs/DAntoniKW18}
Loris D'Antoni, Zachary Kincaid, and Fang Wang.
\newblock A symbolic decision procedure for symbolic alternating finite automata.
\newblock In Sam Staton, editor, {\em Proceedings of the Thirty-Fourth Conference on the Mathematical Foundations of Programming Semantics, {MFPS} 2018, Dalhousie University, Halifax, Canada, June 6-9, 2018}, volume 341 of {\em Electronic Notes in Theoretical Computer Science}, pages 79--99. Elsevier, 2018.
\newblock URL: \url{https://doi.org/10.1016/j.entcs.2018.03.017}, \href {https://doi.org/10.1016/J.ENTCS.2018.03.017} {\path{doi:10.1016/J.ENTCS.2018.03.017}}.

\bibitem{popl/DAntoniV14}
Loris D'Antoni and Margus Veanes.
\newblock Minimization of symbolic automata.
\newblock In Suresh Jagannathan and Peter Sewell, editors, {\em The 41st Annual {ACM} {SIGPLAN-SIGACT} Symposium on Principles of Programming Languages, {POPL} '14, San Diego, CA, USA, January 20-21, 2014}, pages 541--554. {ACM}, 2014.
\newblock \href {https://doi.org/10.1145/2535838.2535849} {\path{doi:10.1145/2535838.2535849}}.

\bibitem{cav/DAntoniV17}
Loris D'Antoni and Margus Veanes.
\newblock The power of symbolic automata and transducers.
\newblock In Rupak Majumdar and Viktor Kuncak, editors, {\em Computer Aided Verification - 29th International Conference, {CAV} 2017, Heidelberg, Germany, July 24-28, 2017, Proceedings, Part {I}}, volume 10426 of {\em Lecture Notes in Computer Science}, pages 47--67. Springer, 2017.
\newblock \href {https://doi.org/10.1007/978-3-319-63387-9\_3} {\path{doi:10.1007/978-3-319-63387-9\_3}}.

\bibitem{DV21}
Loris D'Antoni and Margus Veanes.
\newblock Automata modulo theories.
\newblock {\em Commun. {ACM}}, 64(5):86--95, 2021.
\newblock \href {https://doi.org/10.1145/3419404} {\path{doi:10.1145/3419404}}.

\bibitem{Z3}
Leonardo~Mendon{\c{c}}a de~Moura and Nikolaj Bj{\o}rner.
\newblock {Z3:} an efficient {SMT} solver.
\newblock In C.~R. Ramakrishnan and Jakob Rehof, editors, {\em Tools and Algorithms for the Construction and Analysis of Systems, 14th International Conference, {TACAS} 2008, Held as Part of the Joint European Conferences on Theory and Practice of Software, {ETAPS} 2008, Budapest, Hungary, March 29-April 6, 2008. Proceedings}, volume 4963 of {\em Lecture Notes in Computer Science}, pages 337--340. Springer, 2008.
\newblock \href {https://doi.org/10.1007/978-3-540-78800-3\_24} {\path{doi:10.1007/978-3-540-78800-3\_24}}.

\bibitem{cpp/DoczkalKS13}
Christian Doczkal, Jan{-}Oliver Kaiser, and Gert Smolka.
\newblock A constructive theory of regular languages in coq.
\newblock In Georges Gonthier and Michael Norrish, editors, {\em Certified Programs and Proofs - Third International Conference, {CPP} 2013, Melbourne, VIC, Australia, December 11-13, 2013, Proceedings}, volume 8307 of {\em Lecture Notes in Computer Science}, pages 82--97. Springer, 2013.
\newblock \href {https://doi.org/10.1007/978-3-319-03545-1\_6} {\path{doi:10.1007/978-3-319-03545-1\_6}}.

\bibitem{google-closure-library}
{Google}.
\newblock Google closure library.
\newblock \url{https://developers.google.com/closure/library/}, 2015.
\newblock Accessed: 2015-11.

\bibitem{ostrich}
Matthew Hague, Denghang Hu, Artur Jez, Anthony~W. Lin, Oliver Markgraf, Philipp R{\"{u}}mmer, and Zhilin Wu.
\newblock {OSTRICH2:} solver for complex string constraints.
\newblock In {\em FMCAD}, 2025.

\bibitem{noodler-int}
Vojt\v{e}ch Havlena, Luk\'{a}\v{s} Hol{\'\i}k, Ond\v{r}ej Leng\'{a}l, and Juraj S{\'\i}\v{c}.
\newblock {Cooking String-Integer Conversions with Noodles}.
\newblock In Supratik Chakraborty and Jie-Hong~Roland Jiang, editors, {\em 27th International Conference on Theory and Applications of Satisfiability Testing (SAT 2024)}, volume 305 of {\em Leibniz International Proceedings in Informatics (LIPIcs)}, pages 14:1--14:19, Dagstuhl, Germany, 2024. Schloss Dagstuhl -- Leibniz-Zentrum f{\"u}r Informatik.
\newblock URL: \url{https://drops.dagstuhl.de/entities/document/10.4230/LIPIcs.SAT.2024.14}, \href {https://doi.org/10.4230/LIPIcs.SAT.2024.14} {\path{doi:10.4230/LIPIcs.SAT.2024.14}}.

\bibitem{uss/HooimeijerLMSV11}
Pieter Hooimeijer, Benjamin Livshits, David Molnar, Prateek Saxena, and Margus Veanes.
\newblock Fast and precise sanitizer analysis with {BEK}.
\newblock In {\em 20th {USENIX} Security Symposium, San Francisco, CA, USA, August 8-12, 2011, Proceedings}. {USENIX} Association, 2011.
\newblock URL: \url{http://static.usenix.org/events/sec11/tech/full\_papers/Hooimeijer.pdf}.

\bibitem{pldi/HuD17}
Qinheping Hu and Loris D'Antoni.
\newblock Automatic program inversion using symbolic transducers.
\newblock In Albert Cohen and Martin~T. Vechev, editors, {\em Proceedings of the 38th {ACM} {SIGPLAN} Conference on Programming Language Design and Implementation, {PLDI} 2017, Barcelona, Spain, June 18-23, 2017}, pages 376--389. {ACM}, 2017.
\newblock \href {https://doi.org/10.1145/3062341.3062345} {\path{doi:10.1145/3062341.3062345}}.

\bibitem{cpp/KanLRS22}
Shuanglong Kan, Anthony~Widjaja Lin, Philipp R{\"{u}}mmer, and Micha Schrader.
\newblock Certistr: a certified string solver.
\newblock In Andrei Popescu and Steve Zdancewic, editors, {\em {CPP} '22: 11th {ACM} {SIGPLAN} International Conference on Certified Programs and Proofs, Philadelphia, PA, USA, January 17 - 18, 2022}, pages 210--224. {ACM}, 2022.
\newblock \href {https://doi.org/10.1145/3497775.3503691} {\path{doi:10.1145/3497775.3503691}}.

\bibitem{Kern}
Christoph Kern.
\newblock Securing the tangled web.
\newblock {\em Commun. {ACM}}, 57(9):38--47, 2014.
\newblock URL: \url{http://doi.acm.org/10.1145/2643134}, \href {https://doi.org/10.1145/2643134} {\path{doi:10.1145/2643134}}.

\bibitem{Lammich2014TheCA}
P.~Lammich.
\newblock The {CAVA} automata library.
\newblock {\em Arch. Formal Proofs}, 2014, 2014.

\bibitem{Refine_Monadic-AFP}
Peter Lammich.
\newblock Refinement for monadic programs.
\newblock {\em Archive of Formal Proofs}, January 2012.
\newblock \url{https://isa-afp.org/entries/Refine_Monadic.html}, Formal proof development.

\bibitem{DBLP:conf/itp/Lammich13}
Peter Lammich.
\newblock {Automatic Data Refinement}.
\newblock In Sandrine Blazy, Christine Paulin{-}Mohring, and David Pichardie, editors, {\em Interactive Theorem Proving - 4th International Conference, {ITP} 2013, Rennes, France, July 22-26, 2013. Proceedings}, volume 7998 of {\em Lecture Notes in Computer Science}, pages 84--99. Springer, 2013.
\newblock \href {https://doi.org/10.1007/978-3-642-39634-2\_9} {\path{doi:10.1007/978-3-642-39634-2\_9}}.

\bibitem{DBLP:conf/popl/LinB16}
Anthony~Widjaja Lin and Pablo Barcel{\'{o}}.
\newblock String solving with word equations and transducers: towards a logic for analysing mutation {XSS}.
\newblock In Rastislav Bod{\'{\i}}k and Rupak Majumdar, editors, {\em Proceedings of the 43rd Annual {ACM} {SIGPLAN-SIGACT} Symposium on Principles of Programming Languages, {POPL} 2016, St. Petersburg, FL, USA, January 20 - 22, 2016}, pages 123--136. {ACM}, 2016.
\newblock \href {https://doi.org/10.1145/2837614.2837641} {\path{doi:10.1145/2837614.2837641}}.

\bibitem{afp/LochmannFSTS21}
Alexander Lochmann, Bertram Felgenhauer, Christian Sternagel, Ren{\'{e}} Thiemann, and Thomas Sternagel.
\newblock Regular tree relations.
\newblock {\em Arch. Formal Proofs}, 2021, 2021.
\newblock URL: \url{https://www.isa-afp.org/entries/Regular\_Tree\_Relations.html}.

\bibitem{verified-verifying}
Kevin Lotz, Mitja Kulczynski, Dirk Nowotka, Danny~B{\o}gsted Poulsen, and Anders Schlichtkrull.
\newblock Verified verifying: {SMT-LIB} for strings in isabelle.
\newblock In Benedek Nagy, editor, {\em Implementation and Application of Automata - 27th International Conference, {CIAA} 2023, Famagusta, North Cyprus, September 19-22, 2023, Proceedings}, volume 14151 of {\em Lecture Notes in Computer Science}, pages 206--217. Springer, 2023.
\newblock \href {https://doi.org/10.1007/978-3-031-40247-0\_15} {\path{doi:10.1007/978-3-031-40247-0\_15}}.

\bibitem{z3alpha}
Zhengyang Lu, Stefan Siemer, Piyush Jha, Joel Day, Florin Manea, and Vijay Ganesh.
\newblock Layered and staged monte carlo tree search for smt strategy synthesis.
\newblock In Kate Larson, editor, {\em Proceedings of the Thirty-Third International Joint Conference on Artificial Intelligence, {IJCAI-24}}, pages 1907--1915. International Joint Conferences on Artificial Intelligence Organization, 8 2024.
\newblock Main Track.
\newblock \href {https://doi.org/10.24963/ijcai.2024/211} {\path{doi:10.24963/ijcai.2024/211}}.

\bibitem{Mansur20}
Muhammad~Numair Mansur, Maria Christakis, Valentin W{\"{u}}stholz, and Fuyuan Zhang.
\newblock Detecting critical bugs in {SMT} solvers using blackbox mutational fuzzing.
\newblock In Prem Devanbu, Myra~B. Cohen, and Thomas Zimmermann, editors, {\em {ESEC/FSE} '20: 28th {ACM} Joint European Software Engineering Conference and Symposium on the Foundations of Software Engineering, Virtual Event, USA, November 8-13, 2020}, pages 701--712. {ACM}, 2020.
\newblock \href {https://doi.org/10.1145/3368089.3409763} {\path{doi:10.1145/3368089.3409763}}.

\bibitem{z3str4}
Federico Mora, Murphy Berzish, Mitja Kulczynski, Dirk Nowotka, and Vijay Ganesh.
\newblock Z3str4: A multi-armed string solver.
\newblock In {\em Formal Methods: 24th International Symposium, FM 2021, Virtual Event, November 20–26, 2021, Proceedings}, page 389–406, Berlin, Heidelberg, 2021. Springer-Verlag.
\newblock \href {https://doi.org/10.1007/978-3-030-90870-6_21} {\path{doi:10.1007/978-3-030-90870-6_21}}.

\bibitem{neha}
Neha Rungta.
\newblock A billion {SMT} queries a day (invited paper).
\newblock In Sharon Shoham and Yakir Vizel, editors, {\em Computer Aided Verification - 34th International Conference, {CAV} 2022, Haifa, Israel, August 7-10, 2022, Proceedings, Part {I}}, volume 13371 of {\em Lecture Notes in Computer Science}, pages 3--18. Springer, 2022.
\newblock \href {https://doi.org/10.1007/978-3-031-13185-1\_1} {\path{doi:10.1007/978-3-031-13185-1\_1}}.

\bibitem{Berkeley-JavaScript}
Prateek Saxena, Devdatta Akhawe, Steve Hanna, Feng Mao, Stephen McCamant, and Dawn Song.
\newblock A symbolic execution framework for {JavaScript}.
\newblock In {\em S{\&}P}, pages 513--528, 2010.
\newblock \href {https://doi.org/10.1109/SP.2010.38} {\path{doi:10.1109/SP.2010.38}}.

\bibitem{DBLP:conf/cav/ShiFLTWY20}
Xiaomu Shi, Yu{-}Fu Fu, Jiaxiang Liu, Ming{-}Hsien Tsai, Bow{-}Yaw Wang, and Bo{-}Yin Yang.
\newblock Coqqfbv: {A} scalable certified {SMT} quantifier-free bit-vector solver.
\newblock In Alexandra Silva and K.~Rustan~M. Leino, editors, {\em Computer Aided Verification - 33rd International Conference, {CAV} 2021, Virtual Event, July 20-23, 2021, Proceedings, Part {II}}, volume 12760 of {\em Lecture Notes in Computer Science}, pages 149--171. Springer, 2021.
\newblock \href {https://doi.org/10.1007/978-3-030-81688-9\_7} {\path{doi:10.1007/978-3-030-81688-9\_7}}.

\bibitem{smtlib_benchmarks}
{SMT-LIB}.
\newblock Smt-lib benchmarks.
\newblock \url{https://smt-lib.org/benchmarks.shtml}.
\newblock Accessed: 2025.

\bibitem{sofsem/TammV18}
Hellis Tamm and Margus Veanes.
\newblock Theoretical aspects of symbolic automata.
\newblock In A~Min Tjoa, Ladjel Bellatreche, Stefan Biffl, Jan van Leeuwen, and Jir{\'{\i}} Wiedermann, editors, {\em {SOFSEM} 2018: Theory and Practice of Computer Science - 44th International Conference on Current Trends in Theory and Practice of Computer Science, Krems, Austria, January 29 - February 2, 2018, Proceedings}, volume 10706 of {\em Lecture Notes in Computer Science}, pages 428--441. Springer, 2018.
\newblock \href {https://doi.org/10.1007/978-3-319-73117-9\_30} {\path{doi:10.1007/978-3-319-73117-9\_30}}.

\bibitem{Tuerk-NFA}
Thomas Tuerk.
\newblock {A Formalisation of Finite Automata in Isabelle / HOL}.
\newblock \url{https://www.thomas-tuerk.de/assets/talks/cava.pdf}, 2012.

\bibitem{cvc5}
Stanford University and University of~Iowa.
\newblock cvc5: An efficient open-source automatic theorem prover for smt problems.
\newblock \url{https://cvc5.github.io/}, 2023.
\newblock Accessed: 2023-10-15.

\bibitem{ershov/VeanesB11}
Margus Veanes and Nikolaj~S. Bj{\o}rner.
\newblock Symbolic tree transducers.
\newblock In Edmund~M. Clarke, Irina~B. Virbitskaite, and Andrei Voronkov, editors, {\em Perspectives of Systems Informatics - 8th International Andrei Ershov Memorial Conference, {PSI} 2011, Novosibirsk, Russia, June 27-July 1, 2011, Revised Selected Papers}, volume 7162 of {\em Lecture Notes in Computer Science}, pages 377--393. Springer, 2011.
\newblock \href {https://doi.org/10.1007/978-3-642-29709-0\_32} {\path{doi:10.1007/978-3-642-29709-0\_32}}.

\bibitem{lpar/VeanesBM10}
Margus Veanes, Nikolaj~S. Bj{\o}rner, and Leonardo~Mendon{\c{c}}a de~Moura.
\newblock Symbolic automata constraint solving.
\newblock In Christian~G. Ferm{\"{u}}ller and Andrei Voronkov, editors, {\em Logic for Programming, Artificial Intelligence, and Reasoning - 17th International Conference, LPAR-17, Yogyakarta, Indonesia, October 10-15, 2010. Proceedings}, volume 6397 of {\em Lecture Notes in Computer Science}, pages 640--654. Springer, 2010.
\newblock \href {https://doi.org/10.1007/978-3-642-16242-8\_45} {\path{doi:10.1007/978-3-642-16242-8\_45}}.

\bibitem{VeanesHLMB12Transducer}
Margus Veanes, Pieter Hooimeijer, Benjamin Livshits, David Molnar, and Nikolaj~S. Bj{\o}rner.
\newblock Symbolic finite state transducers: algorithms and applications.
\newblock In John Field and Michael Hicks, editors, {\em Proceedings of the 39th {ACM} {SIGPLAN-SIGACT} Symposium on Principles of Programming Languages, {POPL} 2012, Philadelphia, Pennsylvania, USA, January 22-28, 2012}, pages 137--150. {ACM}, 2012.
\newblock \href {https://doi.org/10.1145/2103656.2103674} {\path{doi:10.1145/2103656.2103674}}.

\bibitem{systematic-transduction}
Joel Weinberger, Prateek Saxena, Devdatta Akhawe, Matthew Finifter, Eui Chul~Richard Shin, and Dawn Song.
\newblock A systematic analysis of {XSS} sanitization in web application frameworks.
\newblock In Vijay Atluri and Claudia D{\'{\i}}az, editors, {\em Computer Security - {ESORICS} 2011 - 16th European Symposium on Research in Computer Security, Leuven, Belgium, September 12-14, 2011. Proceedings}, volume 6879 of {\em Lecture Notes in Computer Science}, pages 150--171. Springer, 2011.
\newblock \href {https://doi.org/10.1007/978-3-642-23822-2\_9} {\path{doi:10.1007/978-3-642-23822-2\_9}}.

\bibitem{osdi/YaseenABCL20}
Nofel Yaseen, Behnaz Arzani, Ryan Beckett, Selim Ciraci, and Vincent Liu.
\newblock Aragog: Scalable runtime verification of shardable networked systems.
\newblock In {\em 14th {USENIX} Symposium on Operating Systems Design and Implementation, {OSDI} 2020, Virtual Event, November 4-6, 2020}, pages 701--718. {USENIX} Association, 2020.
\newblock URL: \url{https://www.usenix.org/conference/osdi20/presentation/yaseen}.

\end{thebibliography}
\end{document}